%% file: lifted-quant-hsp-mfps.tex
\documentclass[final,twoside,11pt]{entics}

\include{preamble}

\FXRegisterAuthor{sm}{asm}{SM}
\FXRegisterAuthor{hu}{ahu}{HU}
\FXRegisterAuthor{jj}{ajj}{JJ}

\begin{document}
\begin{frontmatter}
  \title{Algebraic Reasoning over Relational Structures} 						
  \author{Jan Jurka\thanksref{a}\thanksref{myemail}\thanksref{myfunding}}
  \author{Stefan Milius\thanksref{b}\thanksref{coemail}\thanksref{cofunding}}%
  \author{Henning Urbat\thanksref{b}\thanksref{co2email}\thanksref{co2funding}}%
  \address[a]{Department of Mathematics and Statistics\\ Faculty of Science\\ Masaryk University\\			
    Brno, Czech Republic}  							
  \address[b]{Department of Computer Science\\Friedrich-Alexander-Universit\"at Erlangen-N\"urnberg\\
    Erlangen, Germany}
  \thanks[myemail]{Email: \href{mailto:jurka@math.muni.cz} {\texttt{\normalshape jurka@math.muni.cz}}}
  \thanks[coemail]{Email:  \href{mailto:mail@stefan-milius.eu} {\texttt{\normalshape mail@stefan-milius.eu}}}
  \thanks[co2email]{Email:  \href{mailto:henning.urbat@fau.de} {\texttt{\normalshape henning.urbat@fau.de}}}
  \thanks[myfunding]{Supported by the Grant Agency of the Czech Republic under the grant 22-02964S and by Masaryk University under the grant MUNI/A/1457/2023.}
  \thanks[cofunding]{Supported by the DFG (German Research Foundation) -- project number 517924115.}
  \thanks[co2funding]{Supported by the DFG (German Research Foundation) -- project number 470467389.}

\begin{abstract}
  Many important computational structures involve an intricate interplay between algebraic features (given by operations on the underlying set) and relational features (taking account of notions such as order or distance). This paper investigates algebras over relational structures axiomatized by an infinitary Horn theory, which subsume, for example, partial algebras, various incarnations of ordered algebras, quantitative algebras introduced by Mardare, Panangaden, and Plotkin, and their recent extension to generalized metric spaces and lifted algebraic signatures by Mio, Sarkis, and Vignudelli. To this end, we develop the notion of clustered equation, which is inspired by Mardare et al.'s basic conditional equations in the theory of quantitative algebras, at the level of generality of arbitrary relational structures, and we prove that it is equivalent to an abstract categorical form of equation earlier introduced by Milius and Urbat. Our main results are a family of Birkhoff-type variety theorems (classifying the expressive power of clustered equations) and an exactness theorem (classifying abstract equations by a congruence property).
\end{abstract}
\end{frontmatter}

\section{Introduction}

The axiomatization of data types by operations (constructors) and
equations that these ought to satisfy is an important tool in algebraic
specifications. Accordingly, the theory of models of equational
specifications is a topic of great interest in both mathematics
and computer science. One key result is Birkhoff's seminal variety
theorem, also known as the HSP theorem~\cite{Birkhoff35}. It states that a class of algebras over a
signature $\Sigma$ is a \emph{variety} (i.e.~axiomatizable by equations $s = t$
between $\Sigma$-terms) iff it is closed under
\emph{h}omomorphic images, \emph{s}ubalgebras, and
\emph{p}roducts. Birkhoff also introduced a complete deduction
system for reasoning about equations.

In modern algebraic approaches to the semantics of programming
languages, data types and computational effects, models often involve
an intricate interplay between algebraic features given by operations
on the underlying set and relational features taking account of
notions of order or distance. For example, Bloom~\cite{bloom76} introduced \emph{ordered algebras} (posets equipped with monotone
operations) and established a variety theorem for them along with a
complete deduction system. Here, the role of equations $s=t$ is taken
over by inequations $s\leq t$. Another example is that of \emph{quantitative
  algebras} (metric spaces equipped with nonexpansive operations), introduced by Mardare, Panangaden, and Plotkin
\cite{Mardare16,MardarePP17}, which naturally arise as semantic domains in the theory of
probabilistic computation. In the
quantitative setting, equations $s=_\epsilon t$ are parameterized by a
non-negative real number $\epsilon$ and interpreted as ``$s$ and $t$ have
distance at most $\epsilon$''. Among the main results of the latter work are a variety theorem for
quantitative algebras and a complete deduction system. However, the underlying notion of quantitative
algebra has subsequently turned out to be too restrictive for some applications,
e.g.\ in quantitative term rewriting~\cite{gd23} and machine
learning~\cite{ckpr21}. Therefore, Mio, Sarkis, and
Vignudelli~\cite{msv22,msv23} recently proposed a generalization
of it in two directions: (1)~metric spaces are relaxed to
\emph{generalized metric spaces} where only a fixed but arbitrary
subset of the axioms of a metric is required to hold;
and~(2)~nonexpansivity of operations $\sigma_A\colon A^n\to A$ is not
required w.r.t.~the usual product metric on $A^n$, but w.r.t.~to an
arbitrary functorial choice of a metric on $A^n$ (which might depend
on $\sigma$), specified by a lifting of the set functor $(-)^n$. In
this setting they present a complete deduction system for quantitative
equations. However, a variety theorem classifying the expressive power
of quantitative equations over lifted signatures is still missing.

It is one of the goals and motivations of our paper to fill this
gap. On the way, we will move beyond the metric setting and
investigate algebras and equational theories over general relational
structures axiomatized by an infinitary Horn theory. This not only
highlights that the precise nature of the underlying structures is
largely irrelevant from the perspective of algebraic reasoning, but
also allows us to uniformly cover a number of additional settings of
interest, including partial algebras, various types of ordered
algebras, and quantitative algebras with quantities beyond
non-negative real numbers.

The main new concept developed in our paper is that of a
\emph{$c$-clustered equation} (parametric in a cardinal number~$c$)
for relational algebras. In the special case of quantitative algebras
over metric spaces, this notion has previously appeared in the work of
Milius and Urbat~\cite{mu19}, where it is introduced as a variant of Mardare et
al.'s basic conditional
equations~\mbox{\cite{MardarePP17}}. Informally,
$c$-clustered equations can express properties of algebras that
involve conditions on their variables, e.g.~a conditional commutative
law $x=_{1/2} y\vdash x\bullet y=_{1/4}y\bullet x$ for quantitative
algebras with a binary operation $\bullet$ or
$x\leq y \vdash x\bullet y \leq y\bullet x$ for ordered algebras. The
parameter $c$ controls the level of connectedness between the
variables appearing in the premise. Our main result is the variety
theorem for $c$-varieties (\autoref{thm:variety-thm-quant}), which
states that a class of algebras for a (possibly infinitary) lifted
signature is axiomatizable by $c$-clustered equations iff it is closed
under \emph{$c$-reflexive} quotients, subalgebras, and products. Note
that, unlike in Birkhoff's classical variety theorem, $c$-varieties
need not be closed under all quotients but only under those from a
certain class of quotients depending on the parameter $c$.

Our approach to equations and varieties is based on category
theory. Specifically, we make the key observation that our notion of
$c$-clustered equation is equivalent to an abstract categorical form
of equation earlier introduced by Milius and Urbat~\cite{mu19}. Our
variety theorem for $c$-varieties of relational algebras then emerges
by combining this equivalence with their Abstract Variety Theorem
(recalled in \autoref{sec:abs-var}). The categorical perspective has several
advantages; most importantly, it underlines that $c$-clustered
equations, and the corresponding $c$-reflexive quotients featuring in
the closure properties of varieties, are not an ad-hoc concept but
naturally arise from general principles.  Moreover, it allows us to
isolate the generic parts of the proof of the variety theorem from
arguments specific to the particular setting.

While the main focus of our paper is on the model theory of equations
with relational features, we also provide first steps towards a
complete deduction system for such equations. In this regard, observe
that the completeness of Birkhoff's classical equational
logic~\cite{Birkhoff35} can be derived as an easy consequence of the
exactness property of $\Sigma$-algebras, namely the fact that
quotients of an algebra~$A$ can be represented as congruence relations
on~$A$, which are equivalence relations respected by all the
operations. We establish a corresponding exactness result for algebras
over relational structures, which yields a full characterization of
quotient algebras in terms of suitable relations
(\autoref{T:metricquot}). This turns out to be substantially more
involved than the classical case, making it a result of independent
interest, and we expect that it can serve as a basis for a complete
equational logic in our present setting; see also the discussion in \autoref{S:future}.

Proof details omitted due to space restrictions can be found in the arXiv v2 version~\cite{jmu24_arxiv} of our paper.

%
%

\section{Preliminaries}\label{S:prelim}
We assume that readers are familiar with notions of basic category
theory such as functors, (co)limits, and adjunctions. For a gentle introduction, see Mac Lane's book~\cite{MacLane98}.

Let us briefly recall some categorical terminology we use in the sequel. A
\emph{factorization system} $(\E,\M)$ in a category $\A$ consists of
two classes $\E$ and $\M$ of morphisms in $\A$ such~that
\begin{enumerate}
\item both~$\E$ and $\M$ contain all isomorphisms and are closed under
  composition;
  
\item every morphism $f$ has a factorization $f = m\cdot e$
  where $e\in \E$ and $m\in \M$;
  
\item the \emph{diagonal fill-in} property holds: for every
  commutative square as shown below where $e\in \E$ and $m\in \M$,
  there exists a unique morphism $d$ making both triangles commute.
  \[
    \begin{tikzcd}
      A \ar{r}{e} \ar{d}[swap]{f} & B \ar{d}{g} \ar[dashed]{dl}[swap]{d} \\
      C \ar{r}{m} & D
    \end{tikzcd}
  \] 
\end{enumerate}
The factorization system is \emph{proper} if
all morphisms in~$\E$ are epic and all morphisms in~$\M$ are
monic; in this case, morphisms in $\E$ and $\M$ are denoted by $\epito$ and
$\monoto$, respectively. 
A simple example is the proper factorization system of $\Set$, the category
of sets and functions, given by
$(\E,\M) = (\text{surjective maps},\, \text{injective maps})$. Given a proper factorization system $(\E,\M)$, a
\emph{quotient} of an object~$A$ is represented by a morphism
$e\colon A \epito B$ in $\E$ and a \emph{subobject} by a morphism
$m\colon B \monoto A$ in $\M$. Two quotients $e\colon A\epito B$ and
$e'\colon A\epito B'$ are identified if there exists an isomorphism
$i\colon B\xra{\cong} B'$ such that $e'=i\cdot e$; dually for
subobjects. The category $\A$ is \emph{$\E$-co-well-powered} if for
every object $A$ the class of quotients of $A$, taken up to
isomorphism, forms a small set.

An object $X\in \A$ is called \emph{projective} w.r.t.~a morphism
$e\colon A\to B$ if the induced map
$\A(X,e)=e\cdot(-)\colon \A(X,A)\to \A(X,B)$ between hom-sets is
surjective. In other words, for every \mbox{$h\colon X\to B$}, there
exists a morphism $g\colon X\to A$ such that $h = e\cdot g$.

\section{Algebras over Relational Structures}\label{S:algrel}
In the following we study algebraic structures whose underlying set is
equipped with additional relations, which the operations of the
algebra respect in a user-defined manner.

A \emph{(finitary) relational signature} $\S$ is a set of \emph{relation
  symbols} with associated positive \emph{arity} $\arity(R)\in \Nat_+$
for each $R\in \S$.
An \emph{$\S$-structure} $(A,(R_A)_{R\in \S})$ is given by a set $A$ equipped with an $n$-ary relation $R_A\seq A^n$
for every $n$-ary relation symbol $R\in \S$. We usually just write~$A$
for $(A,(R_A)_{R\in \S})$. A \emph{morphism} $h\colon A\to B$ of
$\S$-structures is a relation-preserving map: for each $n$-ary
$R\in\S$ and $a_1,\ldots,a_n\in A$,
\[
  R_A(a_1,\ldots,a_n)\quad\implies\quad R_B(h(a_1),\ldots, h(a_n)).
\]
Conversely, a map $h\colon A\to B$ is said to \emph{reflect relations} if for each $n$-ary
$R\in\S$ and $a_1,\ldots,a_n\in A$,
\[
  R_A(a_1,\ldots,a_n)\quad\Longleftarrow\quad R_B(h(a_1),\ldots, h(a_n)).
\]
An \emph{embedding} is an injective map $m\colon A \monoto B$ that
both preserves and reflects relations.

We denote the category of $\S$-structures and their
morphisms by $\Str(\S)$, and its forgetful functor by \[U\colon \Str(\S)\to\Set.\]
For every $A\in \Str(\S)$ we write $|A|$ for the cardinality of its underlying set $UA$.
\begin{myremark}\label{rem:str-props}
  \begin{enumerate}
  \item The category $\Str(\S)$ is complete and cocomplete, with limits
    and colimits formed at the level of underlying sets; in particular, $U$ preserves limits and colimits. Specifically:
    \begin{enumerate}
    \item The product $A=\prod_{i\in I} A_i$ is given by the cartesian product
      equipped with the relations defined by
      \[
        R_A((a_{i,1})_{i\in I},\ldots, (a_{i,n})_{i\in I})
        \quad\iff\quad
        \forall i\in I: R_{A_i}(a_{i,1},\ldots, a_{i,n}).
      \]
      
    \item\label{rem:str-props:1b} The coproduct $A=\coprod_{i\in I} A_i$ is given by the disjoint union,
      and $R_A(a_1,\ldots,a_n)$ holds iff $a_1,\ldots,a_n$ lie in the
      same coproduct component $A_i$ and $R_{A_i}(a_1,\ldots,a_n)$.
      
    \item\label{rem:str-props:1c} A diagram $D\colon I\to \Str(\S)$ is
      \emph{$\kappa$-directed}, for a regular cardinal
      number~$\kappa$, if its scheme~$I$ is a $\kappa$-directed poset,
      that is, every subset of $I$ of cardinality less than~$\kappa$
      has an upper bound. A \emph{$\kappa$-directed colimit} is a
      colimit of a $\kappa$-directed diagram. A
      \emph{$\kappa$-directed union of embeddings} is a
      $\kappa$-directed colimit where all connecting morphisms
      $D_i\to D_j$ ($i\leq j$) are embeddings. To form the colimit
      of any $\kappa$-directed diagram $D$, one takes the colimit cocone $c_i\colon UD_i\to C$ ($i\in I$)
      of~$U\!D$ in $\Set$ and equips $C$ with the following relations for each $n$-ary relation symbol $R\in \S$:
      \[
        R_C(x_1,\ldots, x_n)
        \quad \iff\quad
        \exists i\in I.\, \exists y_1,\ldots,y_n\in D_i.\; x_i=c_i(y_i)\,\wedge\,
        R_{D_i}(y_1,\ldots, y_n)
      \]
      In the case of a $\kappa$-directed union, $C$ is the union of the sets~$UD_i$ ($i \in I$) and all
      colimit injections $c_i$ are embeddings.%
      \smnote{I reformulated slightly here; the previous formulation was confusing: were we talking about arbitrary colimits or unions? If the latter why didn't we say `union' before the display in lieu of `colimit'?}
      Moreover, if
      $z_i\colon D_i\to Z$ ($i\in I$) is another cocone over $D$ where
      all $z_i$ are embeddings, then the unique mediating map
      $z\colon C\to Z$ such that $z_i=z\cdot c_i$ for all~$i\in I$ is an
      embedding, too.
    \end{enumerate}
    
  \item The category $\Str(\S)$ has the factorization
    system given by surjective morphisms and embeddings. Accordingly,
    \emph{quotients} and \emph{substructures} of $\S$-structures are
    represented by surjections and embeddings.
  \end{enumerate}
\end{myremark}

In the following we shall consider structures axiomatized by (possibly infinitary) Horn
clauses:
\begin{mydefinition}\label{def:horn-clause}
  An \emph{infinitary Horn clause} over a set $X$ of variables
  is an expression of either of the types
  \begin{align}
    R_i(x_{i,1},\ldots,x_{i,n_i})\;\;(i\in I)\;\;
    &\vdash\;\;
    R(x_1,\ldots,x_n), \label{eq:horn-clause-1}
    \\
    R_i(x_{i,1},\ldots,x_{i,n_i})\;\;(i\in I)\;\;
    &\vdash\;\; x_1=x_2, \label{eq:horn-clause-2} 
  \end{align} 
  where (a)~$I$ is a set, (b)~$x_k, x_{i,k} \in X$ for all indices
  $i,k$, and (c)~$R_i$ ($i\in I$) and $R$ are relation symbols in~$\S$
  with arities $n_i$ and $n$, respectively.
\end{mydefinition}
\begin{mydefinition} Let $A$ be an $\S$-structure.
  \begin{enumerate}
  \item The structure $A$ \emph{satisfies} the clause
    \eqref{eq:horn-clause-1} if for every map\footnote{The map $h$ can be thought of as an assignment of values in $A$ to each variable in $X$.} $h\colon X\to A$,
    \[
      (R_i)_A(h(x_{i,1}),\ldots, h(x_{i,n_i}))\text{ for all $i\in I$}
      \qquad \text{implies} \qquad
      R_A(h(x_1),\ldots, h(x_n)).
    \]
  \item  Similarly, $A$ 
    \emph{satisfies} the clause \eqref{eq:horn-clause-2} if for every map $h\colon X\to A$,
    \[
      (R_i)_A(h(x_{i,1}),\ldots, h(x_{i,n_i}))\text{ for all $i\in I$}
      \qquad \text{implies} \qquad
      h(x_1) = h(x_2).
    \]
  \end{enumerate}
\end{mydefinition}
\begin{mynotation}\label{N:C}
From now on, we fix a relational signature $\S$ and a set $\Ax$ of infinitary Horn clauses over~$\S$.
  %
We denote the full subcategory of structures satisfying all clauses in $\Ax$ by
\[
    \C\hookto \Str(\S).
\]

\end{mynotation}

\begin{mylemma}\label{P:closed}
  The category $\C$ is closed under
  products and
  substructures in $\Str(\S)$.%
\end{mylemma}
%
\begin{myexample}\label{ex:genmetsp}
  Our leading example is that of generalized metric spaces~\cite{msv22}.
  A {\it fuzzy relation} on a set $A$ is a map
  $d \colon A \times A \to [0,1]$. Let $\Ax_{\mathrm{GM}}$ be a fixed
  subset of the following axioms:
  \begin{align*}
    \forall a \in A: d(a,a) &= 0 \tag{Refl} \label{ax:refl} \\
    \forall a,b \in A: d(a,b) &= 0\; \implies\; a = b \tag{Pos} \label{ax:pos}  \\
    \forall a, b \in A: d(a,b) &= d(b,a) \tag{Sym} \label{ax:sym} \\
    \forall a, b, c \in A: d(a, c) &\leq d(a,b) + d(b,c) \tag{Tri} \label{ax:tri} \\ 
    \forall a, b, c \in A: d(a, c) &\leq \max\{d(a,b), d(b,c)\} \tag{Max} \label{ax:max} 
  \end{align*}
  A \emph{generalized metric space} is a set $A$ with a fuzzy
  relation $d_A\colon A\times A\to [0,1]$, subject to the axioms
  in~$\Ax_{\mathrm{GM}}$. A map $h\colon A\to B$ between generalized
  metric spaces is \emph{nonexpansive} if
  $d_B(h(a),h(a'))\leq d_A(a,a')$ for $a,a'\in A$. We let
  $\GMet$ denote the category of generalized metric spaces and nonexpansive
  maps.\footnote{Since $\GMet$ is parametric in the choice of $\Ax_{\mathrm{GM}}$,
    this defines a family of categories.} 

We can regard generalized metric spaces as relational structures as follows. Consider the
  relational signature
  $\S = \{\, =_\epsilon : \epsilon \in [0,1]\, \}$ where
  $\arity(=_\epsilon) = 2$ for each $\epsilon \in [0,1]$. Let
  $\Ax$ be the corresponding\footnote{This means that $\Ax$
    contains~(Up) and~(Arch), and a primed axiom
    appears in $\Ax$ iff the corresponding non-primed axiom appears in
    $\Ax_{\mathrm{GM}}$.} subset of the following Horn clauses, where
  $\epsilon, \epsilon'\in [0,1]$:
  \begin{align*}
    &\;\vdash\; x =_{0} x \tag{Refl$^\prime$} \label{ax:reflc} \\
    x=_{0} y &\; \vdash\; x = y && \tag{Pos$^\prime$} \label{ax:posc} \\
    x =_{\epsilon} y &\; \vdash\; y =_{\epsilon}x && \tag{Sym$^\prime$} \label{ax:symc} \\
    x =_{\epsilon} y,\, y =_{\epsilon'} z &\; \vdash\; x =_{\epsilon + \epsilon'} z && (\epsilon+\epsilon'\leq 1) \tag{Tri$^\prime$} && \label{ax:tric} \\
    x =_{\epsilon} y,\, y =_{\epsilon'} z &\; \vdash\; x =_{\max\{\epsilon, \epsilon'\}} z && \tag{Max$^\prime$} \label{ax:maxc} \\
    x =_{\epsilon}y &\; \vdash\; x =_{\epsilon'} y && (\epsilon<\epsilon') \tag{Up} \label{ax:up} \\
    x =_{\epsilon'} y \;\;(\epsilon'>\epsilon) &\; \vdash\; x =_{\epsilon} y && \tag{Arch} \label{ax:arch}  
  \end{align*}
  An $\S$-structure $(A, (=_\epsilon)_{\epsilon \in [0,1]})$
  satisfying $\Ax$ then gives rise to a generalized metric
  space~$(A,d)$ with the generalized metric defined by
  $d(a,a') := \inf \{ \epsilon : a =_{\epsilon} a' \}$. In the opposite
  direction, a generalized metric space $(A,d)$ defines an
  $\S$-structure $(A,(=_\epsilon)_{\epsilon \in [0,1]})$ where
  $a =_\epsilon a'$ holds iff $d(a,a') \leq \epsilon$. This
  $\S$-structure clearly satisfies $\Ax$. Moreover, these two
  correspondences are mutually inverse:
  \begin{enumerate}
  \item Consider the composite
    $(A,d) \mapsto (A, (=_\epsilon)_{\epsilon \in [0,1]}) \mapsto
    (A,d')$. Then we clearly have
    $d(a,a') = \inf \{ \epsilon : a =_{\epsilon} a'\}$ by the
    definition of $=_\epsilon$. Thus $d(a,a') = d'(a,a')$.%
    %
    
  \item Consider the composite
    $(A, (=_\epsilon)_{\epsilon \in [0,1]}) \mapsto (A,d) \mapsto
    (A, (='_\epsilon)_{\epsilon \in [0,1]})$. Then we have that
    $a =_{\epsilon} a'$ implies $d(a,a') \leq \epsilon$, which
    implies $a='_\epsilon a'$. Conversely, if $a ='_\epsilon a'$,
    then $d(a,a') \leq \epsilon$, and thus for each
    $\epsilon' > \epsilon$ we have $d(a,a') < \epsilon'$. Since $d(a,a')=\inf \{ \delta: a=_\delta a'\}$, there exists
    $\delta < \epsilon'$ such that $a =_{\delta} a'$. By using (Up)
    we see that $a =_{\epsilon'} a'$. This holds for each $\epsilon'>\epsilon$, hence (Arch) yields $a =_{\epsilon} a'$.
  \end{enumerate}
  Furthermore, nonexpansive maps and morphisms of $\C$ are
  clearly in one-to-one correspondence. 
  %
  Consequently, the category $\GMet$ is isomorphic to the category
  $\C$. For the case of (ordinary) metric spaces, where $\Ax_{\mathrm{GM}}$ consists of (Refl), (Pos), (Sym), (Tri), this was already observed by Mardare et al.~\cite{MardarePP17}. 
\end{myexample}
\begin{myexample}
  We mention some further examples of categories of relational structures
  specified by infinitary Horn clauses.
  \begin{enumerate}
  \item The category $\Set$ of sets and functions is specified by the
    empty relational signature and the empty set of axioms.
    
  \item The category $\Pos$ of partially ordered sets (posets) and
    monotone maps is specified by the relational signature $\S$
    consisting of a single binary relation symbol $\leq$ and the
    axioms
    \[
      \vdash\; x\leq x,\qquad		
      x\leq y,\, y\leq z \;\vdash\; x\leq z, \qquad\text{and}\qquad 
      x\leq y,\, y\leq x \;\vdash\; x=y.        
    \]

  \item Let~$L$ be a complete lattice where for every $l\in L$ and
    $P\seq L$ with $l>\bigwedge P$ one has~$l\geq p$ for some
    $p\in P$.  Moreover, let $\S$ be the relational signature consisting of
    binary relation symbols~$=_l$ for all $l\in L$ and consider the
    axioms
    \begin{align*}
      x =_{l}y &\; \vdash\; x =_{l'} y && (l<l') \tag{Up} \\
      x =_{l'} y \;\;(l'>l) &\; \vdash\; x =_{l} y && \tag{Arch}
    \end{align*}
    This specifies the category of
    \emph{$L$-valued relations}: its objects are sets~$X$
    equipped with a map $P\colon X\times X\to L$, and its morphisms
    $(X,P)\to (Y,Q)$ are maps $h\colon X\to Y$ such that
    $Q(h(x),h(y))\le P(x,y)$. Of course, fuzzy relations are the
    special case $L=[0,1]$.
    
  \item A signature of partial operations is a set~$P$ of operation
    symbols $f$ with prescribed arities $\arity(f)\in \Nat$.  A (partial)
    $P$-algebra is given by a set $A$ equipped with a partial
    map $f_A\colon A^{\arity(f)}\to A$ for each $f\in P$. A morphism of partial
    algebras is a (total) map $h\colon A\to B$ such that whenever
    $f_A(x_1,\dots,x_{\arity(f)})$ is defined, then
    $f_B(h(x_1),\dots,h(x_{\arity(f)}))$ is defined and equals
    $h(f_A(x_1,\dots,x_{\arity(f)}))$. The category of partial~$P$-algebras and their morphims is isomorphic to
    the category specified by the relational signature consisting
    of relational symbols $\alpha_f$ of arity $\arity(f)+1$ for all
    $f\in P$ (with $\alpha_f(x_1,\dots, x_{\arity(f)}, y)$ being
    understood as $f(x_1,\dots, x_{\arity(f)})=y$), and the axioms
    $
      \alpha_f(x_1,\dots, x_{\arity(f)}, y), \alpha_f(x_1,\dots, x_{\arity(f)}, z)\;\vdash\; y= z
    $.
  \end{enumerate}
\end{myexample}
Next, we introduce \emph{lifted algebraic signatures} over relational
structures, which extends the corresponding notion by Mio et
al.~\cite{msv22} for the setting of generalized metric spaces.

\begin{mydefinition}\label{D:lifting}
  A functor $G\colon \Str(\S)\to \Str(\S)$ is a \emph{lifting} of
  $F\colon \Set\to \Set$ if the square below commutes:
    \[
      \begin{tikzcd}
        \Str(\S) \ar{d}[swap]{U} \ar{r}{G} & \Str(\S) \ar{d}{U} \\
        \Set \ar{r}{F} & \Set 
      \end{tikzcd}
    \]
\end{mydefinition}

\begin{mydefinition}\label{D:lifted-sig} An \emph{(infinitary) algebraic signature} is a set
  $\Sigma$ of operation symbols $\sigma$ with prescribed arity~$\arity(\sigma)$, a cardinal number.  A \emph{lifted algebraic
    signature} $\hatSigma$ is given by a signature~$\Sigma$ together
  with a lifting $L_\sigma\colon \Str(\S)\to \Str(\S)$ of the functor
  $(-)^n\colon \Set\to\Set$ for every~$n$-ary operation symbol
  $\sigma\in\Sigma$.  Given $A\in \Str(\S)$ we write $L_\sigma(R_A)$
  for the interpretation of the relation symbol $R\in \S$ in the
  structure~$L_\sigma(A)$:
  \[
    L_\sigma(A)
    =
    (A^n, (L_\sigma(R_A))_{R\in \S}).
  \]
\end{mydefinition}
\begin{myassumption}\label{asm:assumption}
  In the remainder of the paper we fix a lifted algebraic signature $\hatSigma$
  with associated lifted functors $L_\sigma$
  ($\sigma\in\Sigma$). We assume that each $L_\sigma$ preserves embeddings. Moreover, we choose a regular cardinal $\kappa$
  such that every operation symbol in $\Sigma$ has arity $<\kappa$;
  hence $\Sigma$ is a \emph{$\kappa$-ary signature}.
\end{myassumption}
\begin{mydefinition}
  A \emph{$\hatSigma$-algebra} is given by an $\S$-structure $A$
  equipped with $n$-ary operations
  \[
    \sigma_A\colon (A^n,(L_\sigma(R_A))_{R\in \S}) \to (A,(R_A)_{R\in\S})
    \qquad \text{in}\quad \Str(\S)
  \]
  for every $n$-ary operation symbol $\sigma\in \Sigma$. A
  \emph{morphism} $h\colon A \to B$ of $\hatSigma$-algebras is a map
  from $A$ to $B$ that is both a $\Str(\S)$-morphism and a $\Sigma$-algebra morphism; the latter means that $h(\sigma_A(a_1,\ldots,a_n))=\sigma_B(h(a_1),\ldots,h(a_n))$ for each $n$-ary operation symbol $\sigma\in \Sigma$ and $a_1,\ldots,a_n$. We let $\Alg{\hatSigma}$ denote the category of
  $\hatSigma$-algebras and their morphisms, and
  $\Alg{\C,\hatSigma}$ the full subcategory of
  \emph{$\hatSigma$-algebras over $\C$}, that is, $\hatSigma$-algebras
  whose underlying $\S$-structure lies in the full subcategory $\C\hookto \Str(\S)$ given by $\Ax$. 
\end{mydefinition}

The use of lifted signatures allows for some flexibility in how the
individual operations of an algebra respect the underlying
relations. This is illustrated by the following examples.
\begin{myexample}\label{E:horn}
  \begin{enumerate}
  \item\label{E:horn:general} For every relational signature $\S$,
    there are two simple choices of a
    lifting~$L_\sigma\colon \Str(\S)\to \Str(\S)$ for an $n$-ary
    operation symbol $\sigma\in\Sigma$:
    \begin{enumerate}
    \item\label{E:horn:general:discrete} The \emph{discrete lifting}
      $L_\sigma^{\mathsf{disc}}$ maps $A\in \Str(\S)$ to $A^n$
      equipped with empty relations. Then the operation
      $\sigma_A\colon A^n\to A$ of a $\hatSigma$-algebra $A$ is
      just an arbitrary homomorphism of $\Sigma$-algebras that
      is not subject to any relational conditions.  
    \item\label{E:horn:general:product} The \emph{product lifting}
      $L_\sigma^{\mathsf{prod}}$ maps $A\in \Str(\S)$ to the product
      structure $A^n$ in $\Str(\S)$. Then the operation $\sigma_A\colon A^n\to A$ is
      relation-preserving w.r.t.~the product structure (\autoref{rem:str-props}).
    \end{enumerate}
    
  \item For the signature $\S=\{=_\epsilon : \epsilon\in [0,1]\,\}$ and
    $\C = \GMet$ we obtain the quantitative~$\hatSigma$-algebras by
    Mio et al.~\cite{msv22}. In \emph{op.~cit.}~and in \cite{msv23} the authors consider
    two non-trivial liftings which are motivated by applications in quantitative term
    rewriting~\cite{gd23} and machine learning~\cite{ckpr21}:
    \begin{enumerate}
    \item\label{E:horn:metric:lipschitz} The \emph{Lipschitz lifting}
      $L_\sigma^{\mathsf{Lip},\alpha}$ for a fixed parameter
      $\alpha\in [1,\infty)$ maps $A\in \Str(\S)$ to $A^n$ equipped
      with the relations $(a_i)_{i<n} =_\epsilon (a_i')_{i<n}$ iff
      $a_i=_{\epsilon/\alpha} a_i'$ for all $i<n$. Then the operation
      $\sigma_A\colon A^n\to A$ of a quantitative $\hatSigma$-algebra
      $A$ is an $\alpha$-Lipschitz map w.r.t.\ the product metric $d$ on $A^n$, which is defined by $d((a_i)_{i < n}, (a'_i)_{i < n}) := \sup_{i < n} d_A(a_i,a'_i)$.
      
    \item\label{E:horn:metric:lukaszyk} The
      \emph{{\L}ukaszyk–Karmowski} lifting
      $L_\sigma^{\mathsf{LK},p}$, for a fixed parameter $p\in (0,1)$
      and a binary operation symbol $\sigma\in \Sigma$, sends
      $A\in \Str(\S)$ to $A^2$ equipped with the relations defined by
      $(a_1,a_2) =_\epsilon (a_1',a_2')$ iff there exist
      $\epsilon_{ij}\in [0,1]$ ($i,j=1,2$) such that
      $a_1 =_{\epsilon_{11}} a_1'$, $a_1
        =_{\epsilon_{12}} a_2'$, $a_2 =_{\epsilon_{21}} a_1'$, $a_2 =_{\epsilon_{22}} a_2'$ and
      $\epsilon =
      p^2\epsilon_{11}+p(1-p)\epsilon_{12}+(1-p)p\epsilon_{21}+(1-p)^2\epsilon_{22}$. Then
      given a quantitative $\hatSigma$-algebra $A$ the operation
      $\sigma_A\colon A^2\to A$ is nonexpansive w.r.t.~the
      {\L}ukaszyk–Karmowski distance~\cite{lukaszyk04}.
    \end{enumerate}
    We note that the above liftings restrict to
    $L_\sigma\colon \GMet\to\GMet$ for suitable choices
    of~$\Ax_{\mathrm{GM}}$. This is the type of lifting studied by Mio
    et al.~\cite{msv22}.
  \item\label{E:horn:poset} For the signature $\S=\{\,\leq\,\}$ and
    $\C = \Pos$ we obtain various notions of \emph{ordered algebras}, i.e.~algebras carried by a poset.
    \begin{enumerate}
    \item The discrete lifting and the product
      lifting correspond to ordered algebras with arbitrary or monotone operations, respectively. The latter are standard ordered algebras studied in the literature~\cite{bloom76}.  
    \item\label{E:horn:discprod} These two liftings admit a common generalization: for a fixed
      subset $S\seq\{1,\ldots,n\}$ and $\sigma\in \Sigma$, let $L_\sigma^S$ be the lifting that sends
      $A\in \Str(\S)$ to $A^n$ with the relation
      $(a_i)_{i<n}\leq (a_i')_{i<n}$ iff $a_i\leq a_i'$ for every $i\in S$.
      An operation $\sigma_A\colon A^n\to A$ is then monotone in precisely the
      coordinates from~$S$.
      
    \item\label{E:horn:lexico} The \emph{lexicographic lifting}
      $L_\sigma^{\mathsf{lex}}$ sends $A\in \Str(\S)$ to $A^n$ with
      $(a_i)_{i<n}\leq (a_i')_{i<n}$ if either
      $(a_i)_{i<n}=(a_i')_{i<n}$, or $a_k\leq a_k'$ for
      $k=\min \{ i<n : a_i\neq a_i' \}$. An operation $\sigma_A\colon
      A^n \to A$ is then monotone w.r.t.~the lexicographic ordering on $A^n$.
    \end{enumerate}
    Furthermore, combinations of the above items are easily
    conceivable, e.g.~we may specify ordered algebras with a monotone operation $\sigma_A\colon A^5\to A$ where the order on $A^5$ is lexicographic in the
    first two coordinates, coordinatewise in the last two, and
    discrete in the third coordinate.
  \end{enumerate}
\end{myexample}
\begin{myremark}\label{rem:functor-alg}
  Since coproducts in $\Str(\S)$ are formed at the level of underlying
  sets, the polynomial endofunctor
  $H_\Sigma = \coprod_{\sigma\in\Sigma} (-)^{\arity(\sigma)}$ on
  $\Set$ associated to the algebraic signature $\Sigma$ lifts to the endofunctor
  $H_{\hatSigma} = \coprod_{\sigma\in \Sigma} L_\sigma$ on $\Str(\S)$,
  and the category $\Alg{\hatSigma}$ is isomorphic to the category of
  algebras for $H_{\hatSigma}$.
\end{myremark}

The next three lemmas establish some simple properties of the category $\Alg{\hatSigma}$.
\begin{mylemma}\label{lem:products}
  The categories $\Alg{\hatSigma}$ and $\Alg{\C,\hatSigma}$ have products.
\end{mylemma}

\begin{proof*}{\bf Proof sketch.}
This is immediate from \autoref{rem:functor-alg} and the well-known
fact that for every endofunctor $H\colon \A\to \A$, the forgetful
functor from the category of $H$-algebras to $\A$ creates limits. More explicitly, 
the product of algebras $A_j$ in $\Alg{\hatSigma}$, $j\in J$, is
  given by their product $\S$-structure $A=\prod_{j\in J} A_j$ (\autoref{rem:str-props}) with operations defined coordinatewise. The product in $\Alg{\C,\hatSigma}$ is formed in the same way, using that~$\C$ is closed under products in $\Str(\S)$ by \autoref{P:closed}.
\end{proof*}

\begin{mylemma}[Homomorphism Theorem]\label{lem:hom-thm}
  Let $e\colon A\epito B$ and $h\colon A\to C$ be $\hatSigma$-algebra
  morphisms with~$e$ surjective. Then $h$ factorizes through $e$ iff the following conditions hold:
  \begin{enumerate}
  \item for every $a,a'\in A$, if $e(a)=e(a')$, then $h(a)=h(a')$;
  \item for every $n$-ary $R\in \S$ and $a_1,\ldots,a_n\in A$, if
      $R_B(e(a_1),\ldots e(a_n))$,
      then $R_C(h(a_1),\ldots,h(a_n))$.
  \end{enumerate}
\end{mylemma}

\begin{proof}
  The `only if' statement is clear. For the `if' statement the
  first condition asserts that there exists a unique map
  $g\colon B\to C$ such that $h=g\cdot e$. Since $e$ is surjective,
  $g$ forms a $\Sigma$-algebra morphism, and by the second
  condition $g$ is also a $\Str(\S)$-morphism. \mbox{Hence, $g$} is a
  $\hatSigma$-algebra morphism.
\end{proof}
\begin{mylemma}\label{lem:fact-system}
  The category $\Alg{\hatSigma}$ has a proper factorization system
  given by $\hatSigma$-algebra morphisms carried by surjections and
  embeddings.
\end{mylemma}
\begin{proof}
  Every $\hatSigma$-algebra morphism $h\colon A\to B$ admits a
  factorization into an surjection followed by an embedding:
  \[
    h = \big(
    \begin{tikzcd}[cramped,column sep = 20]
      A \ar[two heads]{r}{e} & h[A] \ar[tail]{r}{m} & B
    \end{tikzcd}
    \big),
  \] 
  where $h[A]\seq B$ is the image of $h$ (which forms a
  $\hatSigma$-subalgebra of $B$), the morphism $m$ is the inclusion
  map, and the morphism $e$ is the codomain restriction of $h$. For
  the diagonal fill-in property, consider a square as shown below whose outside
  commutes, where $e$ is surjective and~$m$ is an embedding:
  \[
    \begin{tikzcd}
      A \ar[two heads]{r}{e} \ar{d}[swap]{f} & B \ar{d}{g} \ar[dashed]{dl}[swap]{d} \\
      C \ar[tail]{r}{m} & D
    \end{tikzcd}
  \] 
  The homomorphism theorem yields a unique $\hatSigma$-algebra
  morphism $d$ making the upper triangle commutative. Since~$e$ is
  surjective, this implies that the lower triangle also commutes.
\end{proof}
We conclude this section with the construction of free $\hatSigma$-algebras.
\takeout{ 
\begin{myremark}\label{rem:directed-colimits}
  \smnote[inline]{TODO: remark to be rewritten to take into account \autoref{rem:str-props}.1c.} 
  A diagram $D\colon I\to \C$ is \emph{$\kappa$-directed} if its
  scheme $I$ is a $\kappa$-directed poset, that is, every subset of
  $I$ of cardinality $<\kappa$ has an upper bound. A
  \emph{$\kappa$-directed colimit} is a colimit of a $\kappa$-directed
  diagram. If all connecting morphisms $D_i\to D_j$ ($i\leq j$) are
  embeddings, then the colimit of $D$ is formed by taking the colimit
  cocone $c_i\colon D_i\to C$ ($i\in I$) of $UD$ in $\Set$ and
  equipping $C$ with the $\S$-structure defined by
  $R_C(c_i(x_1), \ldots, c_i(x_n))$ iff $R_{C_i}(x_1, \ldots, x_n)$;
  this uniquely defines $R_C$, using that the colimit injections $c_i$
  are jointly surjective as well as directedness of the diagram. It is
  also not difficult to verify that $C$ satisfies the axioms in
  $\Ax$.%
  \smnote{In the metric case we have missed that directedness is
    crucial here and that axioms should hold in
    $C$.}\smnote[inline]{TODO: details to be written; this uses that
    $X$ in \autoref{def:horn-clause} is finite.}  In particular,
  the set $C$ is the union of the sets $C_i$, and every~$c_i$ is an
  embedding. Moreover, if $y_i\colon D_i\to Y$ ($i\in I$) is another
  cocone over $D$ where all $y_i$ are embeddings, then the unique
  mediating map $y\colon C\to Y$ such that $y_i=y\cdot c_i$ for all
  $i\in I$ is an embedding, too. In summary, we have shown that the
  forgetful functor $U\colon \C \to \Set$ preserves and reflects (in
  fact, creates) $\kappa$-directed unions of embeddings.
\end{myremark}}

\begin{mylemma}\label{lem:lift-h-pres-directed-colimits}
  The functor $H_{\hatSigma}$ preserves~$\kappa$-di\-rec\-ted
  unions of embeddings.
\end{mylemma}
\begin{proof}
 Since coproducts in $\Str(\S)$
  preserve embeddings and commute with $\kappa$-directed colimits,
  it suffices to show that each $L_\sigma$ preserves
  $\kappa$-directed unions of embeddings. (Recall that $L_\sigma$ preserves embeddings by \autoref{asm:assumption}.)
  %
  We know that for every $n < \kappa$, the set functor $(-)^n$
  preserves $\kappa$-directed unions because $\kappa$-directed colimits
  in $\Set$ commute with $\kappa$-small limits (in particular
  products with less than $\kappa$ factors). Moreover,
  from~\autoref{rem:str-props}\ref{rem:str-props:1c} we know that given a cocone
  $c_i\colon D_i\monoto C$ ($i\in I$) of embeddings over a
  $\kappa$-directed diagram of embeddings, then $(c_i)_{i\in I}$ is a colimit
  cocone of~$D$ iff $(Uc_i)_{i \in I}$ is a colimit cocone of $U\!D$, where
  $U\colon \Str(\S) \to \Set$ is the forgetful functor. The desired result
  now follows since for every $\sigma\in\Sigma$ of arity $n$ we
  have $U \cdot L_\sigma = (-)^n \cdot U$ (cf.~\autoref{D:lifting})
  and since $(-)^n \cdot U$ preserves $\kappa$-directed unions of
  embeddings.

  \takeout{ 
    Thus, let $D\colon I\to \C$ be a $\kappa$-directed diagram of
    embeddings, with colimit cocone $c_i\colon D_i\to C$ ($i\in
    I$). Then the cocone
    $L_\sigma c_i\colon L_\sigma D_i\to L_\sigma C$ ($i\in I$) has the
    underlying cocone $c_i^n\colon (UD_i)^n\to (UC)^n$ in~$\Set$,
    where $n=\arity(\sigma)$. Since $\kappa$-directed colimits in
    $\Set$ commute with $\kappa$-small limits (in particular
    $\kappa$-fold products), the latter is a colimit cocone of
    $(-)^n\cdot U$. Moreover
    $L_\sigma(R_C)(L_\sigma c_i(x_1), \ldots, L_\sigma c_i(x_n))$ iff
    $L_\sigma(R_{D_i})(x_1,\ldots,x_n)$ for each $n$-ary $R\in \S$,
    $i\in I$ and $x_1\ldots,x_n\in L_\sigma(C_i)$ because each
    $L_\sigma(c_i)$ is an embedding. Hence,
    by~\autoref{rem:str-props}\ref{rem:str-props:1c} the cocone
    $(L_\sigma c_i)_{i\in I}$ is a colimit of $L_\sigma D$ in
    $\Str(\S)$ and therefore in $\C$ by~\autoref{P:closed}.
  }
\end{proof}
\begin{myproposition}\label{prop:free-alg}
 The forgetful functor
  from $\Alg{\hatSigma}$ to $\Str(\S)$ has a left adjoint assigning to
  every $\S$-structure~$X$ the \emph{free $\hatSigma$-algebra}
  $T_{\hatSigma} X$ on $X$. Its underlying $\Sigma$-algebra is the
  free $\Sigma$-algebra~$T_\Sigma X$ on (the underlying set of) $X$,
  carried by the set of all well-founded $\Sigma$-trees over the set~$X$.
\end{myproposition}
In more detail, a \emph{$\Sigma$-tree over $X$} is a possibly infinite ordered tree with nodes labelled in $\Sigma\cup X$, where each node labelled by $\sigma\in \Sigma$ has exactly $\arity(\sigma)$ successors and each $X$-labelled node is a leaf. A $\Sigma$-tree is \emph{well-founded} if it contains no infinite path. If the signature $\Sigma$ is finitary (i.e.\ $\kappa=\omega$), a well-founded $\Sigma$-tree is precisely the (finite) syntax tree of a $\Sigma$-term in the usual sense.

\begin{myremark}\label{rem:initial-chain}
  The proof makes use of a well-known construction, due to
  Ad\'amek~\cite{adamek74}, of free algebras for an endofunctor
  $H\colon \A\to \A$ on a cocomplete category $\A$. Given an
  object~$X$ of~$\A$, the \emph{free-algebra chain of $X$} for $H$ is the ordinal-indexed chain of
  objects $X^\sharp_i$ and connecting morphisms
  $w_{i,j}\colon X^\sharp_i \to X^\sharp_j$ ($i<j$) defined as follows by transfinite
  recursion:
  \[
    \begin{array}{l@{\,}l@{\ }l@{\qquad}l}
      X^\sharp_0 & = & X, \\
      X^\sharp_{j+1} & = & HX^\sharp_j + X & \text{for all ordinals $j$,}\\
      X^\sharp_j &= & \colim_{i<j} X^\sharp_i & \text{for all limit ordinals $j$, and}\\[10pt]
      w_{0,1} & = & \big(X^\sharp_0 = X \xra{\inr} HX^\sharp_0
      + X = X^\sharp_1\big) & \text{is the right-hand coproduct injection,}
      \\
      w_{j+1,k+1} & = &
      \multicolumn{2}{@{}l}{
       \big(X^\sharp_{j+1} = HX^\sharp_j + X
      \xra{H w_{j,k} + \id_X} HX^\sharp_k + X = X^\sharp_{k+1}\big),}\\[5pt]
      \multicolumn{4}{l}{\text{$w_{i,j}$ ($i<j$) is the colimit
          cocone for limit ordinals $j$,}}\\[5pt]
      \multicolumn{4}{l}{\text{$w_{j,j+1}$ ($j$ limit ordinal) is the unique morphism such that $w_{j,j+1}\cdot w_{i+1,j} = w_{i+1,j+1}$ for $i<j$.}}
    \end{array}
    %
    %
  \]
  %
  The free-algebra chain is said to \emph{converge in $\lambda$ steps} if
  $w_{\lambda,\lambda+1}$ is an isomorphism. In this case, the
  inverse $w_{\lambda,\lambda+1}^{-1}\colon HX^\sharp_{\lambda} + X
  \to X^\sharp_\lambda$ yields, by precomposing with the two
  coproduct injections, the structure and universal morphism of a free
  $H$-algebra on $X$. Note that if the functor $H$ preserves the colimit formed
  at some limit step $\lambda$, then the free-algebra chain converges
  in $\lambda$ steps.
\end{myremark}

\begin{proof*}{\bf Proof of \autoref{prop:free-alg}.}
  Since $\Sigma$ is a $\kappa$-ary signature, the polynomial set
  functor $H_\Sigma$ preserves $\kappa$-directed colimits. Hence, for
  every set $X$, the free-algebra chain of $X$ for~$H_\Sigma$
  converges in $\kappa$ steps. Moreover, it is well-known that the
  free $\Sigma$-algebra on $X$ is carried by the set of all
  well-founded $\Sigma$-trees over $X$~(see
  e.g.~\cite[Thm.~2.9]{amm18}). Using that $H_\Sigma$ preserves
  injections and that they are closed under coproducts, an easy
  transfinite induction shows that each connecting map $w_{i,j}$ is
  injective and that the colimits at limit ordinals are unions.

  Similarly, for every $\S$-structure $X$, the free-algebra chain of
  $X$ for $H_{\hatSigma}$ is formed by embeddings (using \autoref{rem:str-props} and that each $L_{\sigma}$ preserves embeddings, see \autoref{asm:assumption}), taking unions at
  limit ordinals. From \autoref{lem:lift-h-pres-directed-colimits} we
  know that the functor~$H_{\hatSigma}$ preserves
  $\kappa$-directed unions of embeddings. Hence, the free-algebra chain of $X$ for
  $H_{\hatSigma}$ converges in~$\kappa$ steps to the free
  $\hatSigma$-algebra on $X$, in symbols: $X^\sharp_\kappa = T_{\hatSigma}
  X$. Moreover, since the forgetful functor $U\colon \Str(\S) \to \Set$
  preserves $\kappa$-directed colimits, we see that it maps the
  free-algebra chain of $X$ for $H_{\hatSigma}$ to the free-algebra chain of $UX$ for
  $H_\Sigma$. In particular, $U(T_{\hatSigma} X)$ is the set of all
  well-founded~$\Sigma$-trees over $X$.
  \takeout{ 
  %
  %
  \smnote[inline]{TODO: reformulate from here.}
  free $\Sigma$-algebra $T_\Sigma X$ generated by the set $X$ is
  constructed via the initial-algebra chain of $X+H_\Sigma$ as the
  $\kappa$-directed union
  $T_\Sigma X=\bigcup_{\lambda<\kappa} T^{(\lambda)}_\Sigma X$ where
  $T^{(\lambda)}_\Sigma X$ is defined via transfinite recursion:
  \begin{align*}
    T^{(0)}_\Sigma X& = \emptyset & \text{(base step)}\\
    T^{(\lambda+1)}_\Sigma X&=X+H_\Sigma(T^{(\lambda)}_\Sigma X), & \text{(successor step)}\\
    \qquad T^{(\lambda)}_\Sigma X &= \bigcup_{\iota<\lambda} T^{(\iota)}_\Sigma X & \text{(limit step)}
  \end{align*}
  From \autoref{lem:lift-h-pres-directed-colimits} we know that the
  lifting $H_{\hatSigma}$ of $H_\Sigma$ preserves $\kappa$-directed
  unions, hence so does $X+H_{\hatSigma}$. It follows that the
  initial-algebra chain of $X+H_{\hatSigma}$ (all of whose connecting
  maps are embeddings) converges after $\kappa$ steps to the free
  $\hatSigma$-algebra $T_{\hatSigma} X$. Moreover, applying the
  forgetful functor $U\colon \C\to\Set$ to that chain yields the
  initial-algebra chain of $X+H_\Sigma$. In particular,
  $U(T_{\hatSigma} X)$ is the free $\Sigma$-algebra $T_\Sigma X$.}
\end{proof*}

\section{Variety Theorems}\label{S:variety}
In this section we establish the variety theorem for $\hatSigma$-algebras
over $\C$, our fixed subcategory of $\Str(\S)$. Rather than stating and proving the theorem from scratch, we will take a more principled approach and present it as an
instance of a general categorical perspective on equations and varieties.

\subsection{Abstract Varieties}\label{sec:abs-var}
We first review the abstract variety theorem by Milius and
Urbat~\cite{mu19}, which characterizes classes of objects in a
category axiomatizable by an abstract notion of equation. We state the theorem
in a slightly simplified form that is sufficient for our
intended application to algebras over relational structures.

Fix a category $\A$ with a proper factorization system $(\E,\M)$, a full subcategory $\A_0\hookto \A$,
and a class $\X$ of objects of $\A$. Informally, we think of $\A$ as
a category of algebraic structures, of $\A_0$ as the subcategory of those algebras over which varieties are formed, and of $\X$ as the class of
term algebras over which equations are formed; see \autoref{ex:classical-birkhoff} below for a simple instantiation. The class $\X$ determines a class of quotients in $\A$ defined by
\begin{equation}\label{eq:ex}
  \E_\X = \{\, e\in \E : \text{every $X\in \X$ is projective w.r.t.~$e$} \,\}.
\end{equation}
An \emph{$\E_\X$-quotient} is a quotient represented by a
morphism in $\E_\X$.
\begin{myremark}\label{rem:exlift}
In order to determine $\E_\X$ in a category $\A$ of algebras with additional structure, it suffices to look at the category of underlying structures. Specifically, suppose that 
\begin{enumerate}
\item the category $\A$ is part of an adjoint situation $F\dashv U\colon \A\to \B$,
\item there is a class $\X'$ objects in $\B$ such that $\X= \{\, FX' \;:\; X'\in \X'\,\}$, and
\item there is a class $\E'$ of morphisms in $\B$ such that $\E= \{\, e\in \A\;:\; Ue\in \E'\, \}$.
\end{enumerate}
In analogy to $\E_\X$, we define the following class $\E'_{\X'}$ of morphisms in $\B$: 
\[\E'_{\X'} \,=\, \{\, e'\in \E' : \text{every $X'\in\X'$ is projective w.r.t. $e'$}\,\}.\] Then the class $\E_\X$ is given by $\E_\X = \{\, e\in \E \; : \; Ue\in \E'_{\X'} \,\}$. Indeed, for all $e\in \E$, one has 
\begin{align*}
e\in \E_\X &\iff \forall X\in \X: \A(X,e) \text{ is surjective}\\
&\iff \forall X'\in \X': \A(FX',e) \text{ is surjective}\\
&\iff  \forall X'\in \X': \B(X',Ue) \text{ is surjective}\\
&\iff Ue\in \E'_{\X'}.
\end{align*} 
\end{myremark}
\begin{mydefinition}\label{def:abstract-variety}
  \begin{enumerate}
  \item\label{abstract-equation} An \emph{abstract equation} is an $\E$-morphism
    $e\colon X\epito E$ where $X\in \X$ and $E\in \A_0$.
    
  \item An object $A\in \A_0$ \emph{satisfies} the abstract equation $e$ if every
    morphism $h\colon X\to A$ factorizes through~$e$, that is,
    $h=g\cdot e$ for some $g\colon E\to A$.
  
  \item Given a class $\mathbb{E}$ of abstract equations, we denote by
    $\V(\mathbb{E})$ the class of objects satisfying all equations in
    $\mathbb{E}$. A class $\V$ of objects of $\A_0$ is an \emph{abstract
      variety} if it is axiomatizable by abstract equations, that is,
    $\V=\V(\mathbb{E})$ for some class $\mathbb{E}$ of equations.
  \end{enumerate}
\end{mydefinition}

The following theorem, which is a special case of a result by Milius and Urbat~\cite[Thm.~3.16]{mu19}, characterizes abstract varieties by
their closure properties:
\begin{mytheorem}[Abstract Variety Theorem]\label{thm:var-theorem}
  Suppose that the category $\A$ is $\E$-co-well-powered and has pro\-ducts, that
  $\A_0\hookto \A$ is closed under products and subobjects, and that every
  object of $\A_0$ is an~$\E_\X$-quotient of some object of $\X$.  Then
  for every class $\V$ of objects of $\A_0$,
 \[ \text{$\V$ is an abstract variety}\qquad \text{iff}\qquad  \text{$\V$
  is closed under $\E_\X$-quotients, subobjects, and products.}\]
\end{mytheorem}

\begin{myremark}
\begin{enumerate}
\item Closure under $\E_\X$-quotients means that for every $\E_\X$-quotient $e\colon A\epito B$ in $\A_0$, if $A\in \V$ then $B\in \V$. In particular, we assume $B\in \A_0$ from the outset.
\item The key condition of \autoref{thm:var-theorem} is the requirement that every object of $\A_0$ is an~$\E_\X$-quotient of some object of $\X$. It captures, on an abstract categorical level, the intuition that the design of a concrete variety theorem needs to strike a balance: if one aims for expressive equations (corresponding to a `large' choice of $\X$), one needs to make sure that the class $\E_\X$ remains rich enough.
\end{enumerate}
\end{myremark}

\begin{myexample}\label{ex:classical-birkhoff}
  The classical Birkhoff Variety Theorem~\cite{Birkhoff35} corresponds to the
  instantiation
\begin{itemize}
  \item $\A = \A_0 =$ $\Sigma$-algebras for a finitary algebraic signature $\Sigma$;
  \item $(\E,\M)$ = (surjective, injective);
  \item $\X$ = all free (term) algebras $T_\Sigma X$, where $X$ is a set of variables.
  \end{itemize}

 \noindent Note that $\E_\X=\E$ (hence varieties are closed under all quotients) and that an abstract equation
  $e\colon T_\Sigma X\epito E$ can be presented via a set of ordinary equations
  between $\Sigma$-terms given by the kernel relation of $e$:
  \[e \;\mathbin{\widehat{=}}\; \{\,s=t : s,t\in T_\Sigma X,\, e(s)=e(t) \,\}.\]
  Indeed, a $\Sigma$-algebra satisfies $e$ iff
  it satisfies the above term equations in the usual sense.
\end{myexample}
\takeout{
\begin{myremark}
  In addition to the class $\X$, in \cite[Thm.~3.16]{mu19} two further
  parameters are present: a subclass $\A_0\seq \A$ in which varieties
  are formed, and a class $\Lambda$ of cardinal numbers (the arities
  of products under which varieties are
  closed). \autoref{thm:var-theorem} amounts to the choice $\A_0=\A$
  and $\Lambda =$ all cardinals.
\end{myremark}
}

\subsection{Varieties of Algebras over Relational Structures}
We now employ the above abstract framework to derive a variety
theorem for algebras over relational structures. The variety theorem
is parametric in a cardinal number $c>1$, which determines the type
of quotients under which varieties are closed. A structure $X\in \Str(\S)$ is called
\emph{$c$-clustered} if it can be expressed as a coproduct
$X=\coprod_{j\in J}{X_j}$ where $|X_j|<c$ for each~$j\in J$. We instantiate the Abstract Variety
\autoref{thm:var-theorem} to the following data:
\begin{itemize}
\item $\A = \Alg{\hatSigma}$ and $\A_0=\Alg{\C,\hatSigma}$ for a
  lifted signature $\hatSigma$ satisfying \autoref{asm:assumption};
  
\item $(\E,\M) = (\text{surjections},\ \text{embeddings})$, cf.\ \autoref{lem:fact-system};
\item $\X =$ all free algebras $T_{\hatSigma} X$ where $X\in \Str(\S)$ is a $c$-clustered structure.
\end{itemize}

We first characterize the class $\E_\X$ as defined in \eqref{eq:ex}. The characterization is based on a generalization of the notion of \emph{$c$-reflexive morphism}
by Mardare et al.~\cite{MardarePP17} from metric spaces to relational structures:
\begin{mydefinition}
  A morphism $e\colon A \to B$ in $\Str(\S)$ is \emph{$c$-reflexive} if
  for every substructure $B_0\seq B$ of cardinality $|B_0|<c$, there exists a
  substructure $A_0\seq A$ such that $e$ restricts to an
  isomorphism in $\Str(\S)$ (i.e.~a bijective embedding) $e_0\colon A_0\xra{\cong} B_0$. If additionally $e$ is surjective, then $e$ is a \emph{$c$-reflexive quotient}. By extension, a quotient in $\Alg{\hatSigma}$ is \emph{$c$-reflexive} if its underlying quotient in $\Str(\S)$ is $c$-reflexive.
\end{mydefinition}
\begin{mylemma}\label{lem:cref}
  The class $\E_\X$ consists of all $c$-reflexive quotients in $\Alg{\hatSigma}$.
\end{mylemma}
\begin{proof}
To prove \autoref{lem:cref}, we apply \autoref{rem:exlift} to the adjunction
  $
  \begin{tikzcd}[cramped, column sep = 20]
    \Alg{\hatSigma}
    \ar[yshift=4pt]{r}
    \arrow[phantom]{r}{\labelstyle\top}
    &
    \Str(\S)
    \ar[yshift=-4pt]{l}
  \end{tikzcd}
  $ and 
\[ \X'= \text{$c$-clustered structures}\qquad\text{and}\qquad \E' = \text{surjective
  $\Str(\S)$-morphisms}.\] It thus suffices to prove the characterization of $\E_\X$ for
  the case where the signature $\Sigma$ is empty; that is, we can
  assume that $\A=\Str(\S)$ and $\X=$ $c$-clustered structures.

  Note that $\X$ is the closure of the class
  $\X_c= \{ X\in \Str(\S) \;:\; \under{X}<c\,\}$ under coproducts.  Since a
  coproduct is projective w.r.t.~some morphism $e$ if all of its
  coproduct components are, we have $\E_\X = \E_{\X_c}$. Hence, it
  suffices to show that, for every surjection $e\colon A\epito B$ in
  $\Str(\S)$,
  \[
    e\in\E_{\X_c} \quad \iff \quad \text{$e$ is $c$-reflexive.}
  \]
  ($\Longrightarrow$) Suppose that $e\in \E_{\X_c}$, and let
  $m\colon B_0\monoto B$ be a substructure of cardinality $<c$. Then
  $B_0\in {\X_c}$, and thus, there exists $g\colon B_0\to A$ such that
  $e\cdot g = m$. Let $A_0 = g[B_0]$. It follows that $e[A_0]=B_0$,
  whence $e\colon A_0\to B_0$ is a surjection that preserves
  relations. It also reflects relations: for every $n$-ary relation
  symbol $R\in \S$ and $g(b_1),\ldots, g(b_n)\in A_0$, we have
  \begin{align*}
    R_{B_0}(e(g(b_1)), \ldots, e(g(b_n)))  \iff\,& R_{B}(m(b_1), \ldots, m(b_n)) \\
    \iff\, & R_{B_0}(b_1,\ldots,b_n) \\
    \implies\, & R_{A_0}(g(b_1),\ldots, g(b_n)).
  \end{align*}
  Moreover, $e$ is injective: for every pair $g(b), g(b')\in A_0$ we have
  \[
    e(g(b)) = e(g(b')) \implies m(b) = m(b') \implies  b = b' \implies g(b) = g(b').
  \]
  Hence $e\colon A_0\to B_0$ is an isomorphism in $\Str(\S)$. This proves that $e$ is $c$-reflexive.

  \medskip\noindent ($\Longleftarrow$) Suppose that $e$ is
  $c$-reflexive, and let $h\colon X\to B$ be a $\Str(\S)$-morphism with
  domain in~$\X_c$, that is, $\under{X}<c$. Then $h[X]\seq B$ has
  cardinality $<c$. Hence, there exists a substructure $A_0\seq A$
  such that~$e$ restricts to an isomorphism
  $e\colon A_0\xra{\cong} h[X]$. For every $x\in X$, let $g(x)$ be the
  unique element of $A_0$ such that $h(x)=e(g(x))$. This defines a
  function $g\colon X\to A$ satisfying $h=e\cdot g$. Moreover, $g$ is
  a $\Str(\S)$-morphism: for each $n$-ary relation symbol $R\in \S$
  and $x_1,\ldots, x_n\in X$,
  \begin{align*}
    R_X(x_1,\ldots,x_n) \implies\,& R_B(h(x_1),\ldots,h(x_n)) \\
    \iff\,&  R_B(e(g(x_1)),\ldots, e(g(x_n))) \\
    \iff\,& R_A(g(x_1),\ldots,g(x_n)).
  \end{align*}
  This proves that $e$ lies in $\E_{\X_c}$.
\end{proof}
\begin{mycorollary}\label{cor:asm-satisfied}
The data $\A$, $\A_0$, $(\E,\M)$, $\X$ satisfies the assumptions of \autoref{thm:var-theorem}.
\end{mycorollary}

\begin{proof} The category $\A=\Alg{\hatSigma}$ has products by \autoref{lem:products},
    and its full subcategory $\A_0=\Alg{\C,\hatSigma}$ is closed under products and
    subalgebras by \autoref{P:closed}. 

Moreover, $\A=\Alg{\hatSigma}$ is co-well-powered w.r.t.~surjective
    morphisms: The collection of all $\hatSigma$-algebras
    carried by a given set $B$ forms a small set, and for every $\hat\Sigma$-algebra $A$ and every
    quotient $e\colon A\epito B$ in $\Alg{\hatSigma}$ one has $|B|\leq |A|$, hence up to
    isomorphism there is only a small set of quotients of $A$.

 Finally, every $\hatSigma$-algebra $A$ is an $\E_\X$-quotient (equivalently, a $c$-reflexive quotient) of
    some free algebra $T_{\hatSigma} X$, where $X$ is a~$c$-clustered
    structure in $\Str(\S)$. Indeed, let $m_j\colon X_j\monoto A$
    ($j\in J$) be the family of all substructures of $A$ such that
    $|X_j|<c$. Then $X=\coprod_{j\in J} X_j$ is $c$-clustered and the
    induced surjection $e=[m_j]_{j\in J}\colon X\epito A$ is $c$-reflexive,
    as is its unique extension $e^\#\colon T_{\hatSigma} X\epito A$ to
    a $\hatSigma$-algebra morphism.
\end{proof}

In the present setting, abstract equations in the sense of 
\autoref{def:abstract-variety}\ref{abstract-equation} are surjective
morphisms $e\colon T_{\hatSigma} X\epito E$ in $\Alg{\hatSigma}$ with
codomain $E\in \Alg{\C,\hatSigma}$, where $X$ is a $c$-clustered
structure. As we shall see in the proof of \autoref{thm:variety-thm-quant}, they translate into the following concrete syntactic notion
of equation:
\begin{mydefinition}\label{def:c-clustered}
  A \emph{$c$-clustered equation} over the set $X$ of variables is an
  expression of either of the types
  \begin{align}
    R_i(x_{i,1},\ldots,x_{i,n_i})\;\;(i\in I)
    \;\; &\vdash  \;\;
    R(t_1,\ldots,t_{n}), \label{eq:c-clustered}
    \\
    R_i(x_{i,1},\ldots,x_{i,n_i})\;\;(i\in I)
    \;\; &\vdash  \;\;
    t_1 = t_2, \label{eq:c-clustered-strict}
  \end{align}
  where (a)~$I$ is a set, (b)~$x_{i,k}\in X$ for all $i,k$,
  (c)~$t_1,\ldots,t_n$ are $\Sigma$-terms over $X$, (d)~$R_i$
  ($i\in I$) and $R$ are relation symbols in $\S$ with respective arities
  $n_i$ and $n$, and (e)~the set $X$ can be expressed as a disjoint union
  $X=\coprod_{j\in J} X_j$ of subsets of cardinality $|X_j|<c$ such that
  for every $i \in I$, the variables $x_{i,1},\ldots,
  x_{i,n_i}$ all lie in the same set~$X_j$. A $c$-clustered equation
  is \emph{unconditional} if $I=\emptyset$.
\end{mydefinition}

\begin{myremark}
\begin{enumerate}
\item The key condition (e) restricts the level of connectedness of the variables. More formally, let the \emph{Gaifman graph} of \eqref{eq:c-clustered}/\eqref{eq:c-clustered-strict} be the undirected graph whose nodes are the variables in $X$ and with an edge between $x,x'\in X$ iff $x,x'$ both occur in $R_i(x_{i,1},\ldots, x_{i,n_i})$ for some $i\in I$. Condition (e) then expresses precisely that the connected components of the Gaifman graph all have cardinality $<c$.
\item The above definition highlights an advantage of our categorical approach: the notion of $c$-clustered equation is guided by the fact that $\X$ consists of free algebras over $c$-clustered structures (and would arguably be rather hard to come up with ad hoc). The particular choice of $\X$ is in turn made to ensure that $\E_\X$ is rich enough to satisfy the categorical assumptions of \autoref{thm:var-theorem}; see also \autoref{rem:basic-vars} below.
\end{enumerate}
\end{myremark}

\begin{mydefinition}
  Let $A$ be a $\hatSigma$-algebra over $\C$.
  \begin{enumerate}
  \item The algebra $A$ \emph{satisfies} the $c$-clustered equation
    \eqref{eq:c-clustered} if for every map $h\colon X\to A$,
    \[
      (R_i)_A(h(x_{i,1}),\ldots, h(x_{i,n_i}))\text{ for all $i\in I$}
      \qquad \text{implies} \qquad
      R_A(\ext h(t_1),\ldots, \ext h(t_n)).
    \]
    Here $\ext h\colon T_\Sigma X\to A$ denotes the unique
    $\Sigma$-algebra morphism extending $h$.
    
  \item Similarly, $A$ \emph{satisfies} the $c$-clustered equation
    \eqref{eq:c-clustered-strict} if for every map $h\colon X\to A$,
    \[
      (R_i)_A(h(x_{i,1}),\ldots, h(x_{i,n_i}))\text{ for all $i\in I$}
      \qquad \text{implies} \qquad
      \ext h(t_1) = \ext h(t_2).
    \]
    
  \item A class of $\hatSigma$-algebras over $\C$ is a \emph{$c$-variety} if it
    is axiomatizable by $c$-clustered equations.
  \end{enumerate}
\end{mydefinition}

\begin{myexample}(Quantitative Algebras)\label{ex:c-clustered:metric}
  For $\C = \GMet$ the $c$-clustered equations are of the form
\[
      x_i=_{\epsilon_i} y_i\;\;(i\in I)\;\; \vdash  \;\; t_1 =_\epsilon t_2 \qquad\text{or}\qquad
      x_i=_{\epsilon_i} y_i\;\;(i\in I)\;\; \vdash  \;\; t_1 = t_2,
 \]
    where (a)~$I$ is a set, (b)~$x_i,y_i\in X$ for all $i\in I$, (c)
    $t_1$, $t_2$ are $\Sigma$-terms over $X$,
    (d)~\mbox{$\epsilon_i, \epsilon \in [0,1]$}, and (e)~the set~$X$ is
    a disjoint union $X=\coprod_{j\in J} X_j$ of subsets of cardinality $|X_j|<c$
    such that for every $i \in I$, the variables
    $x_i$ and $y_i$ lie in the same set~$X_j$.
    For ordinary metric spaces, these equations correspond to the
    $c$-clustered equations introduced by Milius and
    Urbat~\cite{mu19}. Concerning the special case $c=2$, note that (i) $2$-clustered
    equations can only contain trivial premises of the form
    $x=_\epsilon x$, hence are equivalent to unconditional
    equations, and (ii) all quotients are $2$-reflexive. Both (i) and (ii) are not
    true for generalized metric spaces if the axiom \eqref{ax:refl} is
    absent from $\Ax_{\mathrm{GM}}$, in which case $x=_\epsilon x$
    becomes a non-trivial condition.
\end{myexample}

\begin{myremark}\label{rem:c-basic}
In the case of metric spaces, 
{$c$-clustered equations} are closely related to the \emph{$c$-basic equations} introduced by Mardare et al.~\cite{MardarePP17}, where condition (e) in \autoref{ex:c-clustered:metric} is replaced by the simpler (e') $|I|<c$. If $c$ is an infinite regular cardinal, clearly every $c$-basic equation is $c$-clustered (with a single cluster). Conversely, if $\Sigma$ is a $\kappa$-ary signature and $c\geq \kappa$, every $c$-clustered equation is equivalent to a $c$-basic equation~\cite[Rem.~B.17]{mu19a}. However, for $c<\kappa$, $c$-clustered equations are more expressive than $c$-basic equations~\cite[App.~A]{adamek22}.
\end{myremark}

\begin{myexample}(Ordered Algebras)\label{ex:c-clustered-ordered}
For $\C =\Pos$ the $c$-clustered equations are of the form
\[
      x_i \leq y_i\;\;(i\in I)\;\; \vdash  \;\; t_1 \leq t_2 \qquad\text{or}\qquad
      x_i \leq y_i\;\;(i\in I)\;\; \vdash  \;\; t_1 = t_2,
\]
   subject to the conditions (a)--(c) and (e) as in \autoref{ex:c-clustered:metric}.
\end{myexample}

\begin{myremark}
    The $c$-clustered equations for ordered algebras are related to \emph{inequations in context} by Ad\'amek et
    al.~\cite[Def.~3.15]{afms21}. However, their notion of
    signature admit arities being finite posets, which allows to
    encode certain definedness constraints for operations using
    order relations on their arguments. If one restricts arities in
    their setting to finite discrete posets, then inequations in
    context essentially correspond to $\omega$-clustered equations, where~$\Sigma$ is finitary. This is due to the fact that terms formed
    from operations with finite arity only contain finitely many
    variables, and so the index sets $I$ may be chosen to be
    finite. Moreover, algebras in the sense of \emph{op.~cit.}~are
    $\hatSigma$-algebras in our sense where all arities of $\Sigma$
    are finite and where for each operation symbol one chooses the
    discrete lifting
    (\autoref{E:horn}\ref{E:horn:general:discrete}). Ad\'amek et
    al.~also consider \emph{coherent} algebras, where every operation
    is monotone; restricting to finite discrete arities again, these
    algebras correspond to~$\hatSigma$-algebras where all arities are
    finite and where for each operation symbol $\Sigma$ one chooses
    the product lifting
    (\autoref{E:horn}\ref{E:horn:general:product}).
    These~$\hatSigma$-algebras are the classical ordered algebras
    featuring in Bloom's variety theorem~\cite{bloom76}. However,
    varieties in his setting are specified by unconditional
    inequations (without contexts); like in the metric case, these are
    equivalent to the $2$-clustered equations.
\end{myremark}

%

With these preparations at hand, we establish our main result:
\begin{mytheorem}[Variety Theorem]\label{thm:variety-thm-quant}
  A class of $\hatSigma$-algebras over $\C$ is a $c$-variety iff it is closed
  under $c$-reflexive quotients, subalgebras, and products.
\end{mytheorem}  
\begin{proof}
  ($\Longrightarrow$) It suffices to show that for each $c$-clustered
  equation, the class of all $\hatSigma$-al\-ge\-bras satisfying it has
  the required closure properties. We consider equations
  of type~\eqref{eq:c-clustered}; the proof for~\eqref{eq:c-clustered-strict} is analogous.
  \begin{enumerate}
  \item \emph{Closure under products.} Let $A=\prod_j A_j$ be a
    product of $\hatSigma$-algebras over $\C$ where each $A_j$ satisfies
    \eqref{eq:c-clustered}, and suppose that $h\colon X\to A$ is a map
    such that $(R_i)_A(h(x_{i,1}),\ldots h(x_{i,n_i}))$ for all
    $i$. Denote by $\pi_j\colon A\to A_j$ the $j$th
    product projection. Then the map $h_j=\pi_j\cdot h$ satisfies
    $(R_i)_{A_j}(h_j(x_{i,1}),\ldots, h_j(x_{i,n_i}))$ for all $i$
    because~$\pi_j$ is relation-preserving. Since $A_j$ satisfies \eqref{eq:c-clustered} for all $j\in J$, it follows that $R_{A_j}(h_j^\#(t_1),\ldots, h_j^\#(t_n))$, which is equivalent to 
    $R_{A_j}(\pi_j\cdot h^\#(t_1),\ldots, \pi_j\cdot h^\#(t_n))$ for
    all $j\in J$ and hence to $R_A(h^\#(t_1),\ldots, h^\#(t_n))$.
    This proves that $A$ satisfies \eqref{eq:c-clustered}.
    
  \item \emph{Closure under subalgebras.} Suppose that $A$ is a
    $\hatSigma$-algebra over $\C$ satisfying \eqref{eq:c-clustered}
    and that $m\colon B\monoto A$ is a subalgebra. Let
    $h\colon X\to B$ be a map such that
    $(R_i)_B(h(x_{i,1}),\ldots, h(x_{i,n_i}))$ for all~$i$. Then we
    have $(R_i)_A(m\cdot h(x_{i,1}),\ldots, m\cdot h(x_{i,n_i}))$ for
    all $i$ because $m$ is relation-preserving. Since $A$ satisfies~\eqref{eq:c-clustered}, it follows
    that $R_A((m\cdot h)^\#(t_1),\ldots, (m\cdot h)^\#(t_n))$, hence
    $R_A(m\cdot h^\#(t_1),\ldots, m\cdot h^\#(t_n))$, and thus
    $R_B(h^\#(t_1),\ldots, h^\#(t_n))$ because $m$ is an
    embedding. This proves that $B$ satisfies \eqref{eq:c-clustered}.
    
  \item \emph{Closure under $c$-reflexive quotients.} Suppose that
    $e\colon A\epito B$ is a $c$-reflexive quotient of
    $\hatSigma$-algebras over $\C$ and that $A$ satisfies
    \eqref{eq:c-clustered}. Let $h\colon X\to B$ be a map such that
    $(R_i)_B(h(x_{i,1}),\ldots, h(x_{i,n_i}))$ for all $i$. Since \eqref{eq:c-clustered} is $c$-clustered,
    we have $X=\coprod_{j\in J} X_j$ where $|X_j|<c$
    for each $j$, and for each $i\in I$ the variables
    $x_{i,1},\ldots, x_{i,n_i}$ lie in the same set~$X_j$. Since
    $|h[X_j]|\leq |X_j|<c$ and $e$ is $c$-reflexive, the map $e$
    restricts to a $\Str(\S)$-isomorphism $e\colon A_j\xra{\cong} h[X_j]$
    for some $A_j\seq A$. For each $j \in J$ and $x\in X_j$, let
    $g(x)$ be the unique element of $A_j$ such that
    $h(x)=e(g(x))$. This defines a map $g\colon X\to A$ such that
    $h=e\cdot g$. Using that the variables $x_{i,1},\ldots, x_{i,n_i}$
    lie in the same set $X_j$ and $A_j \cong h[X_j]$, it follows from
    $(R_i)_B(h(x_{i,1}),\ldots, h(x_{i,n_i}))$ that
    $(R_i)_A(g(x_{i,1}),\ldots, g(x_{i,n_i}))$ holds. Therefore, since
    this holds for all $i \in I$ and $A$
    satisfies~\eqref{eq:c-clustered}, we have
    $R_A(g^\#(t_1),\ldots, g^\#(t_n))$. Consequently, we have
    $R_B(e\cdot g^\#(t_1), \ldots, e\cdot g^\#(t_n))$, which in turn
    is equivalent to
    $R_B((e\cdot g)^\#(t_1),\ldots, (e\cdot g)^\#(t_n))$ and hence to
    $R_B(h^\#(t_1),\ldots, h^\#(t_n))$. This proves that~$B$ satisfies
    \eqref{eq:c-clustered}.
  \end{enumerate}
  
  \noindent
  ($\Longleftarrow$)
We apply \autoref{thm:var-theorem} to $\A$, $\A_0$, $(\E,\M)$, $\X$ as chosen above (cf.\ \autoref{cor:asm-satisfied}). By the theorem, every class of $\hatSigma$-algebras
  over $\C$ closed under $c$-reflexive quotients, subalgebras, and
  products is
  axiomatizable by abstract equations
  $e\colon T_{\hatSigma} X\epito E$ where $E\in \Alg{\C,\hatSigma}$
  and $X\in \Str(\S)$ is $c$-clustered. Hence, it suffices to show that for every
  such $e$ there exists an expressively equivalent set of
  $c$-clustered equations. We put
  \[
    \Phi = \{\,
    R(x_1,\ldots, x_n) : \text{$R\in \S,\, x_1,\ldots,x_n\in X$ and $R_X(x_1,\ldots,x_n)$}
    \,\}.
  \]
Since the structure $X$ is $c$-clustered, there exist substructures
  $X_j\seq X$ ($j\in J$) of cardinality $|X_j|< c$ such that $X=\coprod_{j\in J} X_j$. From the description of the relations on the coproduct $X$
  (\autoref{rem:str-props}\ref{rem:str-props:1b}) we see that for
  every $R(x_1, \ldots, x_n)$ in $\Phi$ the variables
  $x_1, \ldots, x_n$ lie in the same component $X_j$. Using this we form the
  following set of $c$-clustered equations:
  \begin{equation}\label{eq:cbasic}
    \begin{aligned}
      &~\{\,\Phi \; \vdash\; R(t_1,\ldots,t_{n}) \;:\; \text{$R\in \S,\, t_1,\ldots,t_n \in T_{\hatSigma} X$ and $R_E(e(t_1),\ldots,e(t_n))$}\,\} \\ 
      \cup~&~\{\, \Phi \; \vdash\; t_1 = t_2 \;:\; \text{$t_1,t_2 \in T_{\hatSigma} X$ and $e(t_1) = e(t_2)$}\, \}.
    \end{aligned}
  \end{equation}
  We claim that $e$ and \eqref{eq:cbasic} are expressively equivalent, that is, an algebra $A\in \Alg{\C,\hatSigma}$ satisfies the abstract
  equation $e\colon T_{\hatSigma}X\epito E$ iff it satisfies all the $c$-clustered equations
  in~\eqref{eq:cbasic}. Indeed, we have the following chain of
  equivalences, where the second step follows from the homomorphism
  theorem (\autoref{lem:hom-thm}):
  \begin{align*}
    & \text{$A$ satisfies $e$}  \\
    \iff~
    &\text{for all $h\colon X\to A$ in $\Str(\S)$, the map $\ext h\colon T_{\hatSigma} X\to A$ factorizes through $e\colon T_{\hatSigma} X \epito E$}\\
    \iff~
    & \text{for all $h\colon X\to A$ in $\Str(\S)$,} \\
    & \text{$R_E(e(t_1),\ldots e(t_n))$ implies $R_A(\ext h(t_1),\ldots \ext h(t_n))$, for all $R\in \S$ and $t_1,\ldots,t_n\in T_\Sigma X$}, \\
    & \text{and $e(t_1) = e(t_2)$ implies $\ext h(t_1) = \ext h(t_2)$, for all $t_1,t_2\in T_\Sigma X$} \\
    \iff~
    & \text{for all maps $h\colon {X}\to A$ such that}\\
    & \text{$R_X(x_1,\ldots,x_n)$ implies $R_A(h(x_1),\ldots, h(x_n))$, for all $R\in \S$ and $x_1,\ldots, x_n\in X$,}\\
    & \text{$R_E(e(t_1),\ldots e(t_n))$ implies $R_A(\ext h(t_1),\ldots \ext h(t_n))$, for all $R\in \S$ and $t_1,\ldots,t_n\in T_\Sigma X$}, \\
    & \text{and $e(t_1) = e(t_2)$ implies $\ext h(t_1) = \ext h(t_2)$, for all $t_1,t_2\in T_\Sigma X$} \\
    \iff~
    & \text{for all maps $h\colon {X}\to A$ such that}\\
    & \text{$R_A(h(x_1),\ldots, h(x_n))$ for all $R(x_1,\ldots,x_n)\in \Phi$,}\\
    & \text{$R_E(e(t_1),\ldots e(t_n))$ implies $R_A(\ext h(t_1),\ldots \ext h(t_n))$, for all $R\in \S$ and $t_1,\ldots,t_n\in T_\Sigma X$}, \\
    & \text{and $e(t_1) = e(t_2)$ implies $\ext h(t_1) = \ext h(t_2)$, for all $t_1,t_2\in T_\Sigma X$} \\
    \iff~
    &\text{$A$ satisfies all the c-clustered equations in \eqref{eq:cbasic}}.\tag*{\qed}
  \end{align*}
  \renewcommand{\qed}{}
\end{proof}
As noted in \autoref{ex:c-clustered:metric}, if $\C$ is the category of metric spaces,
every quotient in $\C$ is $2$-reflexive (that is, $\E_\X=\E$), and
$2$-clustered equations are equivalent to unconditional equations. This reasoning carries over to posets, but not to generalized metric spaces, or arbitrary relational structures. However, we can capture unconditional equations and varieties closed under all
quotients via a different choice of $\X$:
\[
  \X \;=\;  \text{all free algebras $T_{\hatSigma} X$ where $X\in \Str(\S)$ is a discrete  structure}.
\]
Here, a structure $X$ is \emph{discrete} if $R_X=\emptyset$ for all
$R\in \S$; hence discrete structures are essentially just sets. It is not difficult to verify that $\E_\X=\E$ and that
abstract equations $e\colon T_{\hatSigma}X\epito E$ are expressively
equivalent to unconditional equations; the reasoning is analogous to the proofs of \autoref{lem:cref} and \autoref{thm:variety-thm-quant}. Consequently, we obtain as a further instance of the Abstract Variety Theorem:

\begin{mytheorem}[Variety Theorem']\label{thm:variety-thm-quant2}
  A class of $\hatSigma$-algebras over $\C$ is axiomatizable by
  unconditional equations iff it is closed under quotients,
  subalgebras, and products.
\end{mytheorem}  

We conclude this section with a discussion of some applications of the variety theorem and of its relation to other approaches in the literature.

\begin{myexample}(Ordered Algebras)
  For $\C = \Pos$, the cardinal number $c=2$, and $\hatSigma$ obtained by taking for every
  operation symbol the product lifting
  (\autoref{E:horn}(1b)), 
  \autoref{thm:variety-thm-quant} yields Bloom's classical variety
  theorem~\cite{bloom76} for ordered algebras. For all other choices of $c$ and $\hatSigma$, \autoref{thm:variety-thm-quant} instantiates to a family of new variety theorems for $c$-varieties of ordered algebras.
\end{myexample}

\begin{myexample}(Quantitative Algebras)\label{rem:basic-vars}
For $\C$ = metric spaces and again the product lifting for every operation symbol, \autoref{thm:variety-thm-quant} yields a refinement of the variety theorem by Mardare et al.~\cite{MardarePP17}: a class of quantitative algebras is axiomatizable by $c$-clustered equations iff it is closed under $c$-reflexive quotients, subalgebras, and products.\footnote{The variety theorem by Mardare et al.~\cite{MardarePP17} works with $c$-basic equations (\autoref{rem:c-basic}) instead of $c$-clustered equations, but its statement is incorrect except for the cases where the two notions are equiexpressive; see~\cite[App.~A]{adamek22} for a counterexample. Note that in our categorical setting, $c$-basic equations correspond to the choice $\X$ = free algebras $T_{\hatSigma} X$ where $|X|<c$. The class $\E_\X$ then still consists of all $c$-reflexive quotients, but the key assumption of \autoref{thm:var-theorem} is no longer satisfied:  not every quantitative algebra is a $c$-reflexive quotient of an algebra in $\X$.} For $\C=\GMet$ and arbitrary liftings, we obtain a family of new variety theorems for generalized quantitative algebras. Let us note that the interesting direction ($\Longleftarrow$) of our proof of \autoref{thm:variety-thm-quant}, which proceeds by relating $c$-reflexive equations to abstract equations, is conceptually rather different from the proof of the quantitative variety theorem in \emph{op.~cit}.
\end{myexample}

\begin{myremark}
Mardare et al.~\cite{MardarePP17} also establish a \emph{quasivariety theorem} for classes of finitary quantitative algebras axiomatizable by finitary conditional equations, i.e.\ expressions
$s_i=_{\epsilon_i} t_i\;\;(i\in I)\;\; \vdash  \;\; s =_\epsilon t$ 
where $I$ is finite and $s_i,t_i$ are arbitrary $\Sigma$-terms (rather than just variables). These classes are characterized by being closed under isomorphism, subalgebras, and subreduced products. Note that the scope of the quantitative quasivariety theorem is orthogonal to the quantitative variety theorem and its generalization to relational structures in the present paper: Unrestricted conditional equations are substantially more expressive than $c$-reflexive equations, while at the same time the quantitative quasivariety theorem neither applies to lifted signatures nor to infinitary operations and equations; in fact, since it is derived from the classical quasivariety theorem in model theory, an infinitary extension would be highly non-trivial.
\end{myremark}

\begin{myremark}
In recent work, Mio et al.\ \cite{msv23} develop an alternative approach to finitary quantitative universal algebra that avoids lifted signatures entirely. Their idea is to consider only quantitative algebras whose operations are arbitrary maps and express any desired nonexpansivity-type conditions via suitable quantitative equations (rather than via nonexpansivity w.r.t.~a lifted signature). For instance, the assertion that an operation
  $\sigma_A\colon A^n\to A$ is $\alpha$-Lipschitz, that is,
  nonexpansive w.r.t.~the Lipschitz lifting
  $L_\sigma^{\mathsf{Lip},\alpha}$ of
  \autoref{E:horn}\ref{E:horn:metric:lipschitz}, may be expressed via
  the equations
  $x_i=_{\epsilon/\alpha} x_i'\;(i<n)\;\vdash\;
  \sigma(x_1,\ldots,x_n) =_{\epsilon} \sigma(x_1',\ldots,x_n')$
  for~$\epsilon\in [0,1]$. Similar equations can be
  given for the other liftings in \autoref{E:horn}. In fact, this approach applies to every finitary lifted signature due to the fact that every lifting of a finitary monad from $\Set$ to $\GMet$ admits a quantitative equational presentation~\cite[Thm~8.11]{msv23}. We conjecture that this result easily extends from generalized metric spaces to arbitrary relational structures with infinitary Horn axioms.

In the context of variety theorems (which are not covered by Mio et al.), using lifted signatures has the advantage of providing an explicit separation between nonexpansivity-type conditions on the operations, and all the other axioms of a quantitative equational theory. This introduces more flexibility in the notion of variety. For example, non-expansivity of a $c$-ary operation can be expressed by a set of $c^+$-clustered equations (where $c^+$ is the successor cardinal of $c$),
but not by $c$-clustered equations. Therefore, in the present setting involving lifted signatures there are $c$-varieties that would not be $c$-varieties otherwise.
\end{myremark}

\begin{myremark}
Rosick\'y~\cite{rosicky24} recently introduced \emph{discrete enriched Lawvere theories} and developed a suitable notion of \emph{clustered equation} at this level of categorical generality. The scope of the variety theorem of \emph{op.~cit.}~is orthogonal to ours: It applies to algebras over locally presentable symmetric monoidal closed categories (which generalize our categories of relational structures), but unlike our present work it involves restrictions on the arities of operations and does not capture lifted signatures.
\end{myremark}
\section{Exactness Property}\label{S:exact}
It is well-known that for every $\Sigma$-algebra $A$, surjective $\Sigma$-algebra morphisms
$e\colon A\epito B$ are in bijective correspondence with congruence
relations on $A$, which are equivalence relations respected by the
operations $\sigma_A\colon A^n\to A$. Here we establish a
corresponding exactness property for~$\hatSigma$-al\-ge\-bras, which
turns out to be more involved and slightly subtle. For notational
simplicity we assume in this section that the signature $\Sigma$ is
finitary; however, all statements easily extend to infinitary
operations.

Recall from \autoref{N:C} that $\C\hookto \Str(\S)$ is the category of $\S$-structures
satisfying the infinitary Horn clauses from $\Ax$. Similarly, let $\C'\hookto \Str(\S)$ denote
the category of $\S$-structures satisfying the infinitary Horn clauses
from $\Ax'$, the set of clauses of $\Ax$ of type
\eqref{eq:horn-clause-1} (that is, clauses of type
\eqref{eq:horn-clause-2} are dropped from $\Ax$). For example, if
$\C$ is the category of metric spaces, then $\C'$ is the category of
pseudometric spaces because the axiom (Pos') of \autoref{ex:genmetsp} is dropped. Note that $\C=\C'$ if~$\Ax$ contains no axioms of
type~\eqref{eq:horn-clause-2}.

%
\begin{mydefinition}\label{def:comp-pair}
  \begin{enumerate}
  \item\label{refining-pmet} Given a $\hatSigma$-algebra $A$ over $\C$ with
    underlying $\S$-structure $(A,(R_A)_{R\in \S})$, a {\it refining
      structure} on $A$ is an $\S$-structure $(R'_A)_{R\in \S}$ with carrier $A$
    satisfying the following properties:
    \begin{enumerate}
    \item\label{refining-pmet1} $(A,(R'_A)_{R\in \S})$ lies in $\C'$;
      
    \item\label{refining-pmet2} $R_A\seq R'_A$ for each $R\in \S$;
      
    \item\label{refining-pmet3} for each $\sigma\in\Sigma$, the
      operation $\sigma_A$ is relation-preserving w.r.t.~the relations $R'_A$ and
      $L_\sigma(R'_A)$: 
      \[
        L_{\sigma}(R'_A)((a_{i,1})_{i < n}, \ldots, (a_{i,m})_{i < n})
        \quad\implies\quad
        R_A'(\sigma_A((a_{i,1})_{i<n}),\ldots, \sigma_A((a_{i,m})_{i<n}) )
        \quad\text{for every $R\in \S$,}
      \]
      where $n$ is a the arity of $\sigma$, $m$ is the arity of $R$, and $a_{i,j}\in A$.
    \end{enumerate}
    
  \item A \emph{congruence} on $A$ is an equivalence relation $\equiv$ on $A$
    such that, for all $\sigma\in \Sigma$ of arity $n$ and all $a_i,
    a_i'\in A$, $i = 1, \ldots, n$, we have
    \[
      a_i\equiv a_i'\;\;(i<n) \quad\implies\quad
      \sigma_A(a_1,\ldots,a_n)\equiv \sigma_A(a_1',\ldots,a_n').
    \]
    
  \item A \emph{compatible pair $((R'_A)_{R\in \S},\equiv)$} on $A$ is
    given by a refining structure $(R'_A)_{R\in \S}$ and a congruence
    $\equiv$ on $A$ satisfying the following conditions:
    \begin{enumerate}
    \item For each $n$-ary $R\in \S$ and $a_i,a_i'\in A$, $i = 1,
      \ldots, n$, we have
      \[
        a_i \equiv a_i'\;\;(i<n)
        \quad\implies\quad
        \big(\, R_A'(a_1,\ldots,a_n) \iff R_A'(a_1',\ldots,a_n')\,\big).
      \]
      
    \item For all axioms of type \eqref{eq:horn-clause-2} in
      $\Ax$ and $h\colon X\to A$,
      \[
        \text{$(R_i')_A(h(x_{i,1}),\ldots, h(x_{i,n_i}))$ for all $i\in I$}
        \qquad \text{implies} \qquad
        h(x_1) \equiv h(x_2).
      \]
    \end{enumerate}
  \end{enumerate}
\end{mydefinition}
\begin{myexample}(Quantitative Algebras)
  For $\C = \GMet$, \autoref{def:comp-pair} can be rephrased as
  follows. A \emph{generalized pseudometric} is a fuzzy relation satisfying all axioms from $\Ax_{\mathrm{GM}}$ except possibly \eqref{ax:pos}. We assume that each lifting $L_\sigma$ restricts to an endofunctor $L_\sigma\colon \GPMet\to \GPMet$ on the category of generalized pseudometric spaces and nonexpansive maps. For each $(A,p)\in \GPMet$ we write $L_\sigma(p)$ for the generalized pseudometric on $L_\sigma(A,p)$.
  \begin{enumerate}
  \item\label{refining-pmetex} Given a $\hatSigma$-algebra
    $(A,d_A)$ over $\GMet$, a fuzzy relation $p$ on $A$ is a {\it refining
      generalized pseudometric} if
    \begin{enumerate}
    \item\label{refining-pmetex1} $p$ is a generalized pseudometric,
      
    \item\label{refining-pmetex2} $p(a,a')\leq d_A(a,a')$ for all
      $a,a'\in A$, and
      
    \item\label{refining-pmetex3} for each $\sigma\in\Sigma$, the
      operation $\sigma_A$ is nonexpansive w.r.t.~$p$ and
      $L_\sigma(p)$:
      \[
        p(\sigma_A((a_i)_{i<n}), \sigma_A((b_i)_{i<n})) \leq
        L_{\sigma}(p)((a_i)_{i < n}, (b_i)_{i < n}).
      \]
    \end{enumerate}
    
  \item A refining generalized pseudometric $p$ and a congruence
    $\equiv$ are \emph{compatible} if
    \[
      a \equiv b\, \wedge\, a' \equiv b' \quad\implies\quad p(a,a') =
      p(b,b')
    \]
    for all $a, a', b, b' \in A$, and furthermore, if
    $\Ax_{\mathrm{GM}}$ contains \eqref{ax:pos}, then
    \[
      p(a,a') = 0 \quad\implies\quad a \equiv a'.
    \]
  \end{enumerate}
  Then a compatible pair ($p,\equiv$) corresponds to a compatible pair
  $(((R'_\epsilon)_A)_{\epsilon \in [0,1]},\equiv)$; the
  correspondence between $p$ and
  $((R'_\epsilon)_A)_{\epsilon \in [0,1]}$ is obtained as in
  \autoref{ex:genmetsp}.
\end{myexample}

Now for each $A\in \Alg{\hatSigma}$ there is an order on compatible pairs on $A$ defined by
\[
  ((R'_A)_{R\in \S},\equiv)\,\leq\, ((R''_A)_{R\in \S},\equiv')
  \qquad\text{iff}\qquad
  \text{$R'_A\seq R''_A$ for all $R\in \S$ and ${\equiv} \seq {\equiv'}$.}
\]
Similarly, quotients of $A$ are ordered by $e\leq e'$ iff
$e'=h\cdot e$ for some $\hatSigma$-algebra morphism~$h$. A
\emph{$\C$-quotient} is a quotient with codomain in
$\Alg{\C,\hatSigma}$. Under the above orders, both compatible pairs
and $\C$-quotients of $A$ form complete lattices.

 We let $\E_{\leftrightarrow}$ denote the class of all quotients in $\Str(\S)$ that both preserve and reflect relations.

\begin{mytheorem}[Exactness]\label{T:metricquot} Suppose that $\hat\Sigma$ is a lifted signature satisfying  $L_\sigma(\E_{\leftrightarrow})\seq \E_{\leftrightarrow}$ for all $\sigma\in\Sigma$.
 Then for $A\in \Alg{\hatSigma}$ the complete lattices of $\C$-quotients of $A$ and compatible pairs on $A$ are isomorphic. 
  %
\end{mytheorem}
Hence, for free algebras $A=T_{\hatSigma} X$ the theorem fully characterizes abstract equations $e\colon T_{\hatSigma} X\epito E$.

\begin{proof*}{\bf Proof sketch.}
  For every $\C$-quotient $e\colon A\epito B$ we obtain the compatible
  pair $((R_e)_{R\in \S},\equiv_e)$ defined by
  \[ R_e(a_1,\ldots,a_n) \iff R_B(e(a_1),\ldots, e(a_n))
    \qquad\text{and}\qquad a\equiv_e a' \iff e(a)=e(a')
  \]
  for each $n$-ary $R\in \S$ and $a_1,\ldots,a_n,a,a' \in A$.  In the
  reverse direction,
  every compatible pair $P=((R'_A)_{R\in \S}, \equiv)$ yields a
  $\C$-quotient by forming the quotient $\hatSigma$-algebra
  $e_P\colon A\epito A/\mathord{\equiv}$, where $A/\mathord{\equiv}$
  is the quotient $\Sigma$-algebra induced by the congruence $\equiv$ 
  with relations defined by
  \[
    R_{A/\mathord{\equiv}}([a_1],\ldots,[a_n])
    \iff
    R_A'(a_1,\ldots,a_n)
  \]
  for each $n$-ary $R\in \S$ and $a_1,\ldots, a_n\in A$. It is not difficult to verify that the two maps
  \[
    e\;\mapsto\; ((R_e)_{R\in \S},\equiv_e)
    \qquad\text{and}\qquad
    P=((R_A')_{R\in \S}, \equiv) \;\mapsto\; e_P
  \]
  are monotone and mutually inverse.
\end{proof*}


\takeout{
\section{Equational Logic}
In this section we give a sound and complete deduction system for premise-free equations over relational structures. Similar to the variety theorem, the completeness that emerges is presented as a special case of a completeness result for abstract equations.

\subsection{Abstract Equational Logic}
We consider again the abstract setting of \autoref{thm:var-theorem}, and we assume that $\X$ is chosen to be the class of $\E$-projective objects, so that $\E=\E_\X$.

\begin{mydefinition}
We say that an equation $e\colon X\epito E$ \emph{semantically entails} the equation $e'\colon X'\epito E'$, denoted $e\models e'$, if every object $A\in \A$ satisfying $e$ also satisfies $e'$. 
\end{mydefinition}
To get a sound and complete deduction system for equations, we need the notion of an equational theory:

\begin{mydefinition}
An \emph{abstract equational theory} is given by a family $(e_X\colon X\epito E)_{X\in \X}$ such that for every $h\colon X'\to X$ that the morphism $e_X\cdot h$ factorizes through $e_{X'}$.
\[
\begin{tikzcd}
X' \ar{r}{h} \ar[two heads]{d}[swap]{e'} & X \ar[two heads]{d}{e} \\
E' \ar[dashed]{r}{\exists} & E
\end{tikzcd}
\]
\end{mydefinition}
Informally, if we think of $X'$ and $X$ as term algebras, an equational theory is a family of equations closed under all substitution instances.

\begin{myremark}
Equational theories are ordered by $(e_X)_{X\in \X}\seq (e'_X)_{X\in \X}$ iff $e_X\leq e_{X}'$ for all $X\in \X$. For every equation $e\colon X\epito E$, there exists a least theory $(\ol{e}_Y)_{Y\in \X}$ such that $e\leq e_Y$, called the \emph{substitution closure} of $e$.
\end{myremark}

We consider the following proof rules
{ \flushleft

\begin{tabular}{ l l }
 (Weakening) & $e\vdash e'$ for all equations $e,e'$ such that $e\leq e'$. \\
 (Substitution) &  $e\vdash \ol{e}_Y'$ for all equations $e$ and all $Y\in \X$.
\end{tabular}
}
  
\smallskip\noindent 
By extension, we write $e\vdash e'$ if $e'$ is provable from $e$ using a finite chain of applications of the above rules. This yields a sound and complete proof system:

\begin{mytheorem} Under the assumptions of \autoref{thm:var-theorem}, we have for all equations $e,e'$:
\[ e\models e' \qquad\text{iff}\qquad e\vdash e'. \]
\end{mytheorem}

\subsection{Equational Logic over Relational Structures}
We now instantiate the above to the setting of algebras over relational structures:
\begin{itemize}
\item  $\A = \Alg{\C,\hatSigma}$; 
\item $(\E,\M) = (\text{surjections},\ \text{embeddings})$, cf.\ \autoref{lem:fact-system};
\item $\X =$ all free algebras $T_{\hatSigma} \hatX$ where $X\in \Str(\C)$ is discrete.
\end{itemize}
Recall from \autoref{thm:variety-thm-quant2} and its preceding paragraph that $\E=\E_\X$, as required for the abstract equational logic. We consider the following proof rules, where $X$ ranges over sets, regarded as discrete structures in $\Str(\S)$.

\renewcommand{\arraystretch}{.0}
\begin{table}[!h]
\begin{tabularx}{\textwidth}{lX}
 (Ax1) & For each axiom \eqref{eq:horn-clause-1} in $\Ax$ over variables $V$ and each map $h\colon V\to T_{\hatSigma} \hatX$:\\
 &   \[R_i(h(x_{i,1}),\ldots,h(x_{i,n_i}))\;\;(i\in I)\;\;
    \vdash\;\;
    R(h(x_1),\ldots,h(x_n)).\] \\
 (Ax2) & For each axiom \eqref{eq:horn-clause-2} in $\Ax$ over variables $V$ and each map $h\colon V\to T_{\hatSigma} \hatX$:\\
& \[R_i(h(x_{i,1}),\ldots,h(x_{i,n_i}))\;\;(i\in I)\;\;
    \vdash\;\;
    h(x_1) = h(x_n).\] \\
(Lift) &  For each $\C'$-structure $(X',(S_{X'})_{S\in \S})$, $R\in \S$ and each $\C'$-morphism $h\colon X'\to T_{\hatSigma}\hatX$, the following rule for all $x_{i,k}\in X'$ such that $L_{\sigma}(R_{X'})((x_{i,1})_{i < n}, \ldots, (x_{i,m})_{i < n})$: \\
& \[ \{ S(x_{1},\ldots, x_{m_S}) : S\in \S,\, S_X(x_1,\ldots, x_{m_S}) \} \;\;\vdash\;\; R(\sigma((h(x_{i,1}))_{i<n}),\ldots, \sigma_A((h(x_{i,m}))_{i<n}) ) \]\\
(EqLog) & For all $s,t,u,s_i,t_i\in T_{\hatSigma} \hatX$ and $n$-ary $\sigma\in\Sigma$:
{\begin{align*}
\;\;& \vdash\;\; s=s \\
 s=t \;\;&\vdash\;\; t=s \\
    s=t,\, t=u \;\;& \vdash\;\;  s=u\\
 s_i=t_i\;(i<n) \;\;&\vdash\;\; \sigma(s_1\ldots,s_n)=\sigma(t_1,\ldots,t_n) 
\end{align*}} \\
(Comp) & For all $R\in \S$ and $s_i,t_i\in T_{\hatSigma}\hat X$:
\[ R(s_1,\ldots,s_n),\, s_i=t_i\, (i<n) \;\;\vdash\;\; R(t_1,\ldots,t_n) \] \\
(Subst) & For every $h\colon X\to T_{\hatSigma}\widehat{Y}$ in $\C$, $R\in \S$ and $s,t,t_i\in T_{\hatSigma}\hatX$:
{\begin{align*}
 R(t_1,\ldots,t_n) \;\;&\vdash\;\; R(h^\#(t_1),\ldots, h^\#(t_n)) \\
 s=t\;\;&\vdash\;\; h^\#(s) = h^\#(t) 
\end{align*}}
\end{tabularx}
\end{table}
Note that the rules (Ax1), (Ax2), (Lift), (EqLog) and (Comp) reflect exactly the conditions on compatible pairs which characterize abstract equations (\autoref{T:metricquot}). Similarly, the rule (Subst) reflects the substitution closure property of abstract equational theories.
}

\section{Conclusions and Future Work}\label{S:future}
We have investigated clustered algebraic equations over relational
structures, which generalizes and unifies a number of related notions
that naturally appear in algebraic reasoning over metric spaces or
posets. Our key insight is that this notion is actually an instance
of abstract morphic equations in a categorical framework and that the
characterization of its expressive power can be presented as an
instance of an abstract Birkhoff-type variety theorem. Apart from
simplifying proofs, the generality of the categorical approach highlights
the clear separation between algebraic and relational aspects in
equational reasoning, which often remains implicit when
algebras over specific structures (such as metric spaces) are
considered.

A natural next step is to derive a complete deduction system for
equations with relational features. This should be achievable in a
systematic manner much like in our approach to the variety theorem by
combining the abstract completeness theorem by Milius and
Urbat~\cite[Thm.~4.4]{mu19} with our exactness theorem for relational
algebras (\autoref{T:metricquot}). We expect a tight connection\takeout{, and
possibly a common roof,} to existing completeness results for
generalized quantitative algebras by Mio et al.~\cite{msv22,msv23} and
for algebras over infinitary Horn structures by Ford et
al.~\cite{fms21-2}. \takeout{The latter work considers (unlifted) algebraic
signatures with non-discrete arities of operations; its
dimension of generality is somewhat orthogonal to ours.}

A further direction is to relate our work to the recent investigation
of monads over metric spaces~\cite{adamek22,adv23,rosicky21} and
posets~\cite{adv22,afms21}. In all these works, notions of equational
theories are characterized by properties of their corresponding
free-algebra monads. In our setting, characterizing the monads on the
category $\C$ corresponding to $c$-varieties remains an open problem,
which we expect to be quite challenging in general.

Another potential direction is to investigate whether clustered algebraic equations over relational
structures can be expressed alternatively via Lawvere theories with arities \cite{mel10}.

\bibliographystyle{./entics}
\bibliography{refs}

\end{document}

%% file: preamble.tex
\usepackage{amssymb,amsxtra,mathrsfs,mathtools,bm,stmaryrd}
\usepackage{xspace}
\usepackage{dsfont}
\usepackage{tikz}
\usetikzlibrary{arrows,decorations.markings,decorations.pathreplacing,arrows.meta,fit,calc,positioning}
\usepackage{ifthen}
\usepackage[notref,notcite]{showkeys}
\usepackage[noadjust]{cite}
\usepackage{rotating}
\usepackage{tikz-cd}
\usepackage{blkarray}
\usepackage{adjustbox}
\usepackage{calc}
\usepackage{aliascnt}

\usepackage[inline]{enumitem}
\setlist[enumerate,1]{label=(\arabic*),font=\normalfont,align=left,leftmargin=0pt,labelindent=0pt,listparindent=\parindent,labelwidth=0pt,itemindent=!,topsep=3pt,parsep=0pt,itemsep=3pt,start=1}
\setlist[enumerate,2]{label=(\alph*),font=\normalfont,labelindent=*,leftmargin=*,start=1, topsep=5pt, itemsep=3pt}
\setlist[itemize]{labelindent=*,leftmargin=*,topsep=5pt,itemsep=3pt}
\setlist[description]{labelindent=*,leftmargin=*,itemindent=-1 em}

\tikzcdset{scale cd/.style={every label/.append style={scale=#1},
    cells={nodes={scale=#1}}}}

\usepackage{accents}

\usepackage{enticsmacro}
\usepackage{graphicx}
\usepackage[babel]{csquotes}
\newcommand{\Nat}{{\mathbb N}}

\def\conf{MFPS 2024}
\def\myconf{mfps40} 	
\volume{4}			
\def\volu{4}			
\def\papno{13}			
\def\mydoi{14598}%
\def\lastname{Jurka, Milius, and Urbat} 
\def\runtitle{Algebraic Reasoning over Relational Structures} 
\def\copynames{J. Jurka, S. Milius, and H. Urbat}    

\newaliascnt{mythm}{thm}
\newtheorem{mythm}[mythm]{Theorem}
\aliascntresetthe{mythm}

\newenvironment{mytheorem}{\vspace{-\lastskip}\par \addvspace{.6pc plus
    .2pc minus .1pc}\begin{mythm}}{\end{mythm}\par\addvspace{.6pc
    plus .2pc minus .1pc}}

\newaliascnt{mylem}{thm}
\newtheorem{mylem}[mylem]{Lemma}
\aliascntresetthe{mylem}

\newenvironment{mylemma}{\vspace{-\lastskip}\par \addvspace{.6pc plus
    .2pc minus .1pc}\begin{mylem}}{\end{mylem}\par\addvspace{.6pc
    plus .2pc minus .1pc}}

\newaliascnt{mycor}{thm}
\newtheorem{mycor}[mycor]{Corollary}
\aliascntresetthe{mycor}

\newenvironment{mycorollary}{\vspace{-\lastskip}\par \addvspace{.6pc plus
    .2pc minus .1pc}\begin{mycor}}{\end{mycor}\par\addvspace{.6pc
    plus .2pc minus .1pc}}

\newaliascnt{myprop}{thm}
\newtheorem{myprop}[myprop]{Proposition}
\aliascntresetthe{myprop}

\newenvironment{myproposition}{\vspace{-\lastskip}\par \addvspace{.6pc plus
    .2pc minus .1pc}\begin{myprop}}{\end{myprop}\par\addvspace{.6pc
    plus .2pc minus .1pc}}

\newaliascnt{mydef}{thm}
\newtheorem{mydef}[mydef]{Definition}
\aliascntresetthe{mydef}

\newenvironment{mydefinition}{\vspace{-\lastskip}\par \addvspace{.6pc plus
    .2pc minus .1pc}\begin{mydef}\rm}{\end{mydef}\par\addvspace{.6pc
    plus .2pc minus .1pc}}

\newaliascnt{mynot}{thm}
\newtheorem{mynot}[mynot]{Notation}
\aliascntresetthe{mynot}

\newenvironment{mynotation}{\vspace{-\lastskip}\par \addvspace{.6pc plus
    .2pc minus .1pc}\begin{mynot}\rm}{\end{mynot}\par\addvspace{.6pc
    plus .2pc minus .1pc}}

\newaliascnt{myasm}{thm}
\newtheorem{myasm}[myasm]{Assumption}
\aliascntresetthe{myasm}

\newenvironment{myassumption}{\vspace{-\lastskip}\par \addvspace{.6pc plus
    .2pc minus .1pc}\begin{myasm}\rm}{\end{myasm}\par\addvspace{.6pc
    plus .2pc minus .1pc}}

\newaliascnt{myrem}{thm}
\newtheorem{myrem}[myrem]{Remark}
\aliascntresetthe{myrem}

\newenvironment{myremark}{\vspace{-\lastskip}\par \addvspace{.6pc plus
    .2pc minus .1pc}\begin{myrem}\rm}{\end{myrem}\par\addvspace{.6pc
    plus .2pc minus .1pc}}

\newaliascnt{myexp}{thm}
\newtheorem{myexp}[myexp]{Example}
\aliascntresetthe{myexp}

\newenvironment{myexample}{\vspace{-\lastskip}\par \addvspace{.6pc plus
    .2pc minus .1pc}\begin{myexp}\rm}{\end{myexp}\par\addvspace{.6pc
    plus .2pc minus .1pc}}

\newcommand{\resetCurThmBraces}{%
\gdef\curThmBraceOpen{(}%
\gdef\curThmBraceClose{)}}
\resetCurThmBraces
\newcommand{\removeThmBraces}{%
\gdef\curThmBraceOpen{}%
\gdef\curThmBraceClose{}}
\resetCurThmBraces

\usepackage{etoolbox}
\patchcmd{\thmhead}{(#3)}{\curThmBraceOpen #3\curThmBraceClose }{}{}

\usepackage{wrapfig}

\hypersetup{final,bookmarks}

\usepackage{seqsplit}
\usepackage{xstring}
\newcommand{\defaultshowkeysformat}[1]{%
\StrSubstitute{#1}{ }{\textvisiblespace}[\TEMP]%
\parbox[t]{\marginparwidth}{\raggedright\normalfont\small\ttfamily\(\{\){\color{red!50!black}\expandafter\seqsplit\expandafter{\TEMP}}\(\}\)}%
}

\renewcommand*\showkeyslabelformat[1]{%
\noexpandarg%
\defaultshowkeysformat{#1}%
}

\usepackage[footnote,marginclue,nomargin]{fixme}

\newcommand{\V}{\mathcal{V}}

\newcommand{\seq}{\subseteq}
\newcommand{\xra}[1]{\xrightarrow{~#1~}}

\newcommand{\ol}{\overline}

\newcommand{\op}[1]{\operatorname{\mathsf{#1}}}
\newcommand{\id}{\op{id}}

\newcommand{\ext}[1]{{#1}^\sharp}

\newcommand{\inr}{\op{inr}}

\newcommand{\under}[1]{|#1|}

\newcommand{\cat}[1]{\mathscr{#1}}
\def\A{\cat A}
\def\B{\cat B}
\def\C{\cat C}

\newcommand{\E}{\mathcal{E}}

\newcommand{\Set}{\mathsf{Set}}
\newcommand{\Pos}{\mathsf{Pos}}

\newcommand{\arity}{\mathsf{ar}}

\newcommand{\Ax}{\mathsf{Ax}}

\newcommand{\M}{\mathcal{M}}

\newcommand{\X}{\mathscr{X}}

\newcommand{\Alg}[1]{\mathbf{Alg}(#1)}

\renewcommand{\epsilon}{\varepsilon}

\newcommand{\Str}{\mathop{\mathsf{Str}}}
\newcommand{\colim}{\mathop{\mathsf{colim}}}

\renewcommand{\S}{\mathscr{S}}

\newcommand{\GMet}{\mathsf{GMet}}
\newcommand{\GPMet}{\mathsf{GPMet}}

\renewcommand{\phi}{\varphi}

\newcommand{\hookto}{\hookrightarrow}

\newcommand{\epito}{\twoheadrightarrow}

\newcommand{\monoto}{\rightarrowtail}

\DeclareMathSymbol{\mathinvertedexclamationmark}{\mathord}{operators}{'074}
\DeclareMathSymbol{\mathexclamationmark}{\mathord}{operators}{'041}
\makeatletter
\newcommand{\raisedmathinvertedexclamationmark}{%
  \mathord{\mathpalette\raised@mathinvertedexclamationmark\relax}%
}
\newcommand{\raised@mathinvertedexclamationmark}[2]{%
  \raisebox{\depth}{$\m@th#1\mathinvertedexclamationmark$}%
}
\makeatother

\usepackage[all,2cell]{xy}
\UseAllTwocells
\SelectTips{cm}{}
\newdir{ >}{{}*!/-5pt/@{>}}
\newdir{ (}{{}*!/-10pt/\dir^{(}}
\newdir{ )}{{}*!/-10pt/\dir_{(}}

\newcommand{\takeout}[1]{\empty}

\tikzset{shiftarr/.style={
        rounded corners,%
        to path={--([#1]\tikztostart.center)
                     -- ([#1]\tikztotarget.center) \tikztonodes
                     -- (\tikztotarget)},
}}

\tikzset{shiftarrr/.style={
        rounded corners,%
        to path={-- ([#1]\tikztostart.center)
                    |- (\tikztotarget)  \tikztonodes},
}}

\tikzset{roundcornerarr/.style={
        rounded corners,%
        to path={--([#1]\tikztostart.south)
                     |- (\tikztotarget) \tikztonodes},
}}

\tikzset{roundcornerarrr/.style={
        rounded corners,%
        to path={ -| (\tikztotarget) \tikztonodes},
}}

\tikzset{roundcornerarrrr/.style={
        rounded corners,%
        to path={ |- (\tikztotarget) \tikztonodes},
}}

\newcommand{\hatSigma}{\widehat\Sigma}
\newcommand{\hatX}{\widehat{X}}

\ExplSyntaxOn
\NewDocumentCommand{\makecycle}{om}{
	\ensuremath{ \left(~\guest_print_list:nn { #2 } {~{\ }~}~\right)\IfValueT{#1}{^{#1}}}
}
\NewDocumentCommand{\maketuple}{om}{
	\ensuremath{ \left(~\guest_print_list:nn { #2 } {~{,\ }~}~\right)\IfValueT{#1}{^{#1}}}
}
\NewDocumentCommand{\makemonad}{om}{
	\ensuremath{ \left\langle~\guest_print_list:nn { #2 } {~{,\ }~}~\right\rangle\IfValueT{#1}{^{#1}}}
}

\seq_new:N \l_guest_list_seq
\cs_new_protected:Nn \guest_print_list:nn
{
	\seq_set_from_clist:Nn \l_guest_list_seq { #1 }
	\seq_use:Nn \l_guest_list_seq { #2 }
}
\ExplSyntaxOff

\numberwithin{equation}{section}